\documentclass[review]{elsarticle}

\usepackage{lineno}
\usepackage[colorlinks,hyperindex,breaklinks]{hyperref}
\AtBeginDocument{%
  \hypersetup{
    colorlinks=false,
    linkcolor=black,
    filecolor=blue,      
    urlcolor=blue,
    citecolor=blue}}

\usepackage{changepage}
\usepackage{amsmath}
\usepackage{xcolor}
\usepackage{geometry}
\usepackage{listings}
\usepackage{multirow}
\usepackage{multicol}
\usepackage{tablefootnote}
\usepackage[symbol]{footmisc}
\usepackage{dblfloatfix}
\usepackage{caption}
\usepackage{subcaption}
\usepackage[page,toc,titletoc,title]{appendix}
\usepackage{tocloft}

\definecolor{darkgray}{rgb}{.4,.4,.4}

\lstdefinelanguage{JavaScript}{
  keywords={typeof, new, true, false, catch, function, return, null, catch, switch, var, if, in, while, do, else, case, break},
  ndkeywords={class, export, boolean, throw, implements, import, this},
  sensitive=false,
  comment=[l]{//},
  morecomment=[s]{/*}{*/},
  morestring=[b]',
  morestring=[b]"
}

\modulolinenumbers[5]

\journal{Astronomy and Computing}

\biboptions{authoryear}

\newcommand{\LCDM}{{$\Lambda$CDM} }
\newcommand{\PyCosmo}{{\texttt{PyCosmo}}}
\newcommand{\class}{{\texttt{CLASS}}}

\newcommand{\sympy}{{\texttt{SymPy}} }

\newcommand{\C}{{\texttt{C}} }
\newcommand{\CCpp}{{\texttt{C/C++}} }

\newcommand{\sympytoc}{{\texttt{sympy2c}} }

\begin{document}
\begin{frontmatter}

\title{Symbolic Implementation of Extensions of the \linebreak \PyCosmo\ Boltzmann Solver}

\author[a]{Beatrice Moser,%
\footnote[2]{E-mail: \href{moserb@phys.ethz.ch}{moserb@phys.ethz.ch}}}
\author[a]{Christiane S. Lorenz}
\author[b]{Uwe Schmitt}
\author[a]{Alexandre R\'{e}fr\'{e}gier}
\author[a]{Janis Fluri}
\author[a]{Raphael Sgier}
\author[a]{Federica Tarsitano}
\author[c]{Lavinia Heisenberg}
\address[a]{Institute for Particle Physics and Astrophysics, ETH Zürich, Wolfgang-Pauli-Strasse 27, CH-8093 Zürich, Switzerland}
\address[b]{Scientific IT Services, ETH Zürich, Binzm\"{u}hlestrasse 130,
CH-8092 Zürich, Switzerland}
\address[c]{Institute for Theoretical Physics, ETH Zürich, Wolfgang-Pauli-Strasse 27, CH-8093 Zürich, Switzerland}

\begin{abstract}
\PyCosmo\ is a \texttt{Python}-based framework for the fast computation of cosmological model predictions. One of its core features is the symbolic representation of the  Einstein-Boltzmann system of equations. Efficient \CCpp code is generated from the \sympy symbolic expressions making use of the \sympytoc package. This enables easy extensions of the equation system for the implementation of new cosmological models. We illustrate this with three extensions of the \PyCosmo\ Boltzmann solver to include a dark energy component with a constant equation of state, massive neutrinos and a radiation streaming approximation. 
We describe the \PyCosmo\ framework, highlighting new features, and the symbolic implementation of the new models. We compare the \PyCosmo\ predictions for the $\Lambda$CDM model extensions with \class, both in terms of accuracy and computational speed. We find a good agreement, to better than $0.1\%$ when using high-precision settings and a comparable computational speed. Links to the Python Package Index (PyPI) page of the code release and to the \texttt{PyCosmo Hub}, an online platform where the package is installed, are available at: \href{https://cosmology.ethz.ch/research/software-lab/PyCosmo.html}{https://cosmology.ethz.ch/research/software-lab/PyCosmo.html}.
\end{abstract}

\begin{keyword}
cosmology,  dark energy, cosmological neutrinos, symbolic and algebraic manipulation, solvers
\end{keyword}

\end{frontmatter}
\newpage
\addtocontents{toc}{\protect\enlargethispage{\baselineskip}}
\tableofcontents

\section{Introduction}
\label{sec:intro}
Our understanding of the Universe relies on the possibility to predict cosmological observables from theoretical principles. One of the key theoretical predictions is the evolution over time of the linear order perturbations of the constituents of the Universe, captured by the Einstein-Boltzmann equations (see, e.g.,~\cite{ma}, \cite{dodelson}). The system of ordinary differential equations, due to its complexity and the coupling of the fields, needs to be solved numerically (see, e.g., ~\cite{Ghosh_2017}). For this purpose, several codes have been developed since the release of the pivotal Boltzmann code \texttt{COSMICS} \citep{Bertschinger_1995}, closely followed in time by \texttt{CMBFAST} \citep{Seljak_1996}, later ported to \texttt{C++} with the name \texttt{CMBEASY} \citep{Doran_2005}. The currently maintained Boltzmann solvers are \texttt{CAMB} \citep{camb}, \class\ \citep{class} and \PyCosmo\  \citep{pycosmo}. Both for \texttt{CAMB} and \class\, several codes have been written to include extensions beyond $\Lambda$CDM, for example \texttt{hi\_class} \citep{Zumalacarregui:2016pph} and \texttt{EFTCAMB} \citep{Hu:2013twa} for modified gravity theories, \texttt{CLASS\_EDE} \citep{Hill:2020osr} for early dark energy and \texttt{CLASSgal} \citep{Dio_2013} to include general relativistic effects in the computation of galaxy number counts. \PyCosmo\ was introduced by \cite{pycosmo} as a novel \texttt{Python} library that uses symbolic representation of equations for generating efficient \CCpp code. The framework includes both a Boltzmann solver as well as prediction tools for the computation of cosmological observables with several different fitting functions and approximations \citep{Tarsitano_2020}. With these tools, \PyCosmo\ offers similar utilities as, e.g., the Core Cosmology Library \texttt{CCL} \citep{LSSTDarkEnergyScience:2018yem} developed by the Dark Energy Science Collaboration (DESC). 

An important feature of \PyCosmo\ is the possibility to easily implement model extensions in symbolic form in the code, while taking advantage of the computational speed of the generated \CCpp code. This feature has been improved by rewriting and refactoring  the related code as a  new \sympytoc package presented in \cite{uwe}, which expands the idea of generating fast \CCpp code from symbolic representations of equations. Both \sympytoc and \PyCosmo\ are publicly available in  the  \texttt{Python}  Package  Index  (PyPI). In this work, we illustrate how the \PyCosmo\ Boltzmann solver can be extended thanks to the symbolic framework by implementing several extensions. We introduce two extensions of the Standard Model of Cosmology: a constant dark energy equation of state and massive neutrinos. We also include a radiation streaming approximation (RSA) for photons and massless relics, which approximates the evolution of radiation at late times and includes reionisation, following the treatment in \cite{Blas_2011} implemented in \class. This approximation speeds up the code by reducing the number of equations in the ODE system and avoiding the reflection of power caused by the truncation of the multipole expansion of the  radiation equations. 

We begin by giving an overview of the  new features of the \PyCosmo\ framework in Section \ref{sec:pycosmo2}, where we discuss also the usage of the code and the precision settings used to compare with \class. We then present the equations of the  models, implementation details and code comparisons with \class, both in terms of agreement and performance. In Section \ref{sec:wcdm} we present the implementation of the constant dark energy equation of state, since it is a minimal modification of the Boltzmann system of equations for $\Lambda$CDM. We describe the inclusion of massive neutrinos, treated as non-interacting and non relativistic relics, in Section \ref{sec:massive_nu}. In Section \ref{sec:RSA} we present the Radiation Streaming Approximation, which requires \sympytoc to handle a switch between two different equation systems and is then applied to all models. In Section \ref{sec:discuss} we discuss the results obtained by benchmarking the speed of the computations and comparing the numerical results to \class.  We conclude in Section \ref{sec:conclusion}. This work heavily relies on the Boltzmann equations presented in \cite{ma}. To translate the \PyCosmo\ notation to the Ma-Bertschinger and \class\ notation, we refer the reader to Appendix~\ref{app:notation}. Appendix~\ref{app:eqns} presents the Einstein-Boltzmann system of ODE in \PyCosmo\ notation, using $\ln{a}$ as the independent variable. The adiabatic initial conditions for $\Lambda$CDM and all the other implemented models are shown in Appendix~\ref{app:ini}. We report in Appendix~\ref{app:constants} the parameters of \PyCosmo\ and \class\ that have been kept constant throughout the paper. In Appendix~\ref{app:pk}, we provide a self contained summary of the computation of the total matter power spectrum, including general relativistic corrections, and using the $\mathcal{A}_s$ normalisation parameter. 

\section{PyCosmo framework}
\label{sec:pycosmo2}
\subsection{\CCpp code generation}
\label{sec:codegen}
We reimplemented and improved the \CCpp code generation related parts of  previous versions of \PyCosmo\ \citep{pycosmo} as a separate \texttt{Python} package named \sympytoc which we describe in detail in \cite{uwe}. \sympytoc translates symbolic representations of expressions and ordinary differential equations to \CCpp code and compiles this code as a \texttt{Python} extension module. 

\sympytoc replaces the Backward-Differentiation-Formula (BDF) solver from the previous version of \PyCosmo\ by  the established and robust Livermore Solver for Ordinary Differential Equations (LSODA) solver \citep{lsoda} for improved step-size control and error diagnostics. LSODA detects stiff and nonstiff time domains automatically and switches  between the nonstiff Adams method and the stiff BDF method. The BDF method solves a linear system derived from the Jacobian matrix of the differential equations at each time step. This affects runtime significantly for large systems. \sympytoc leverages the symbolic form of the ODE and generates code to solve such systems efficiently by avoiding unnecessary  computations based on the known sparsity structure of the involved Jacobian matrix.

To solve such linear systems \sympytoc  unrolls loops occurring in used LU factorization with partial pivoting (LUP) algorithm during code generation. 
This procedure depends on predetermined row permutations of the system, and the generated code includes checks for whether the considered permutation is appropriate for ensuring numerical accuracy. When solving the ODE, a new, not yet considered, permutation might arise. In this case, the solver delegates to a fall-back general LUP solver and records the new permutation. The result is a valid result but with a sub-optimal computation time. In this case, the warning message \texttt{"there are new permutations pending, you might want to recompile"} will be displayed together with the command necessary to recompile. Running the code-generator again will then also create optimized code for the newly recorded permutation(s), so that future runs of the solver will benefit from this. This approach starts with the identity permutation and could require several steps of solving the Boltzmann equations followed by code generation and compilation to achieve optimal performance. In our experiments not more than one such iteration is needed. 

A large ODE systems can result in \CCpp functions with millions of lines of code which challenge the compiler and can cause long compilation times and high memory consumption, especially during the optimization phase of the compiler. To mitigate this, \sympytoc can split the original matrix into smaller blocks and then generate code to implement blocked Gaussian elimination using Schur-complements. This affects the generated \CCpp code by creating more but significantly shorter functions and thus supports the optimization step of the compiler. Another benefit of this approach is that runtime is improved by reducing the number of cache misses on the CPU.

Depending on the size and sparsity structure of the system, this code generation and compilation step can take seconds up to 30 minutes or even more. \PyCosmo\ and \sympytoc\ use caching strategies that consider previously generated code so that cached solvers are available within fractions of a second.

\subsection{Usage}

The equations for $\Lambda$CDM and the extended models are implemented symbolically in \PyCosmo, both with and without radiation streaming approximation, in the \texttt{CosmologyCore\_}\textit{model}\texttt{.py} and \texttt{CosmologyCore\_}\textit{model}\texttt{\_rsa.py} files. Supported models are currently \texttt{"lcdm"}, \texttt{"wcdm"} or \texttt{"mnulcdm"}.

The method \texttt{PyCosmo.build} initializes an instance of the \texttt{Cosmo} class for subsequent computations. \texttt{Cosmo} is the
class that manages most of the functionalities of PyCosmo and on which all the other classes rely. \texttt{PyCosmo.build}  requires the name of the model as well as all parameters which influence the code generation and compilation step. The argument \texttt{rsa} enables or disables the RSA and \texttt{l\_max} specifies the maximum moment for truncating the photons and massless neutrinos hierarchies. The \texttt{"mnulcdm"} model also accepts parameters \texttt{l\_max\_mnu} and \texttt{mnu\_relerr} which we describe later. Furthermore, parameters controlling the compiler, such as the optimization flag \texttt{-O}\textit{n} and the \texttt{splits} to use to reduce memory consumption and compilation time, can be specified when calling \texttt{PyCosmo.build}. All the parameters that can be passed to \texttt{build} are specified in Table~\ref{tab:prec}.

Parameters which do not affect code generation, such as cosmological parameters, precision settings, parameters specific for approximations and physical constants can be set or modified using the \texttt{Cosmo.set} method. Each cosmological model is equipped with a default set of such parameters, contained in a \texttt{default\_}\textit{model}\texttt{.ini} file. 
Listing \ref{ls:cosmo} demonstrates how to create a cosmology and change parameters.

\begin{lstlisting}[caption={Creation of a $\Lambda$CDM   cosmology using RSA and truncating photons and massless neutrino hierarchies at \texttt{l\_max} = 50. We override the default value of $\Omega_m$ by $0.28$ and print a report of all parameter values.},captionpos=b, label=ls:cosmo]
import PyCosmo
cosmo = PyCosmo.build("lcdm", l_max=50, rsa=True)
cosmo.set(omega_m=0.28)
cosmo.print_params()
\end{lstlisting}

One parameter that is particularly relevant is \texttt{pk\_type}, since it allows the user to switch between the Boltzmann solver (\texttt{pk\_type} = \texttt{"boltz"}) and the approximations (\texttt{pk\_type} = \texttt{"EH"} for the fitting function by Eisenstein and Hu \citep{Eisenstein_1998} and \texttt{pk\_type} = \texttt{"BBKS"} for the BBKS polynomial fitting function \citep{Peacock_1997}). In this work we will always set \texttt{pk\_type = "boltz"}. Other cosmological and precision parameters which are kept fixed throughout the paper are reported in \ref{app:constants}. Tutorials for the computation of cosmological observables and the usage of the Boltzmann solver can be found on the \texttt{PyCosmo Hub} (see \cite{Tarsitano_2020}), a public platform hosting the current version of \PyCosmo, along with \class, \texttt{CCL}, and \texttt{iCosmo} \citep{iCosmo}, an IDL predecessor of \texttt{PyCosmo}. The link to the \texttt{PyCosmo Hub} can be found at \href{https://cosmology.ethz.ch/research/software-lab/PyCosmo.html}{https://cosmology.ethz.ch/research/software-lab/PyCosmo.html}.

\subsection{Code comparisons setup}
\label{sec:precision}
In order to validate the newly introduced models, we carry out detailed comparisons with \class \footnote{We use \class\ v3.1.0 throughout the paper, through the \texttt{Python} wrapper \texttt{classy}.}. We evaluate the accuracy in terms of relative difference: $$\mathrm{Relative\ difference(\PyCosmo,\class)} = \frac{X_{\PyCosmo} - X_{\class}}{X_{\PyCosmo}} $$ where $X$ is the cosmological observable we want to compare, for example the total matter power spectrum. Since this is a function of the wavenumber $k$, we compare it visually by plotting as a function of $k$ or, when specified, we look at the maximum relative difference in a $k$ interval. 

Comparing the two codes implies carefully setting the cosmological parameters and the precision settings. In the case of \class, we use the precision files shipped with the code: \texttt{cl\_permille.pre} for fast and accurate computation and \texttt{pk\_ref.pre} for high precision. \texttt{cl\_permille.pre} guarantees a precision of $0.1\%$ up to $k = 1 h \mathrm{Mpc^{-1}}$ for the matter power spectrum, and \texttt{pk\_ref.pre} a precision of $0.001\%$ on scales $k < 50 h \mathrm{Mpc^{-1}}$ \citep{lesgourgues2011cosmic}. Both \class\ precision settings use the Radiation Streaming Approximation, described in detail in Section \ref{sec:RSA}. They also include the Tight Coupling Approximation and Ultra-relativistic Fluid Approximation (see \cite{Blas_2011}), whereas only \texttt{cl\_permille.pre} uses the fluid approximation for massive neutrinos (presented in \cite{Lesgourgues_2011}). The main parameters controlling precision in \PyCosmo\ are the \texttt{l\_max} parameter which defines the truncation of the multipole hierarchy for radiation fields (same for photons and massless neutrinos) and the \texttt{boltzmann\_rtol} and \texttt{boltzmann\_atol} parameters defining the relative and absolute tolerance of the LSODA ODE solver. Massive neutrinos also add two important precision parameters which will be described more in detail in section \ref{sec:massive_nu}: \texttt{mnu\_relerr} controlling the number of momenta $q$ used for the massive neutrino integrals and \texttt{l\_max\_mnu}, controlling the truncation of the multipole expansion for massive neutrinos. We summarise all the precision parameters available in \PyCosmo\ in Table~\ref{tab:prec}.
\begin{table}[h]
    \centering
    \begin{tabular}{lccc}
    \hline
    Parameter & Description & Method & Deafult\\
    \hline
    \texttt{rsa} & Switch for the RSA  & \texttt{build} & \texttt{False} \\
    \texttt{compilation\_flags} & GCC compiler's optimization flag & \texttt{build} & ``-O3" \\
    \texttt{splits} & Splittings of the ODE system & \texttt{build} & None \\
    \vspace{-7pt}
    \texttt{reorder} & Whether to reorder the ODE system to speed up & \texttt{build} & \texttt{True} \\
     & code generation  when using \texttt{splits}& & \\
    \texttt{l\_max} & Hierarchy truncation of the relativistic relics & \texttt{build} & 20\\
    \texttt{l\_max\_mnu} & Hierarchy truncation of massive neutrinos & \texttt{build} & 20 \\
    \texttt{mnu\_relerr} & Relative error for massive neutrinos integral & \texttt{build} & 1e-5\\
    \hline
    \hline
    \vspace{-7pt}
    \texttt{sec\_factor} & Safety factor for permuting rows  & \texttt{set} & 10\\
    &  in the LUP decomposition within LSODA &  & \\
    
    \texttt{boltzmann\_rtol} & Relative tolerance of the LSODA solver & \texttt{set} & 1e-5 \\
    \texttt{boltzmann\_atol} & Absolute tolerance of the LSODA solver & \texttt{set} & 1e-5\\
    \texttt{boltzmann\_max\_bdf\_order} & Maximum order used by the BDF integrator & \texttt{set} & 5 \\
    \texttt{boltzmann\_max\_iter} & Max number of iterations of the LSODA solver& \texttt{set} & 2000000\\
    \hline
    \end{tabular}
    \caption{Description of the precision parameters available in the \PyCosmo\ Boltzmann solver, along with the method they are passed to and their default values.}
    \label{tab:prec}
\end{table}

The two precision settings that we use in \PyCosmo\ when comparing respectively to \class\ \texttt{cl\_permille.pre} and \texttt{pk\_ref.pre} are:
 \begin{itemize}
    \item \textit{speed}: \texttt{l\_max} = $17$, \texttt{l\_max\_mnu} = $17$, \texttt{rtol} = $10^{-5}$, \texttt{atol} = $10^{-5}$, \texttt{mnu\_relerr} = $10^{-5}$
    \item \textit{precision}: \texttt{l\_max} = $50$, \texttt{l\_max\_mnu} = $50$, \texttt{rtol} = $10^{-6}$ and \texttt{atol} =$10^{-6}$, \texttt{mnu\_relerr} = $10^{-6}$.
\end{itemize}
When using the RSA in \PyCosmo, we set the RSA trigger parameters (detailed in Section \ref{sec:RSA}) to:
\begin{itemize}
    \item \textit{speed}: \texttt{rsa\_trigger\_taudot\_eta} = 5, \texttt{rsa\_trigger\_k\_eta} = 45 
    \item \textit{precision}: \texttt{rsa\_trigger\_taudot\_eta} = 100, \texttt{rsa\_trigger\_k\_eta} = 240 
\end{itemize}
which match the equivalent \class\ parameters in  \texttt{cl\_permille.pre} and \texttt{pk\_ref.pre}. All the other precision parameters have default values, as in Table~\ref{tab:prec}.
We do not attempt to exactly match the precision parameters in the two packages, since \class\ includes a number of approximations that are not available in \PyCosmo. It is possible to switch off most of the approximations but this would imply losing the precision guarantees of the default precision files.

The cosmological parameters that remain constant throughout the paper are listed in Appendix~\ref{app:constants}, whereas the matter energy density $\Omega_m$, the number of massless neutrinos $N_{\nu}$, the number of massive neutrinos $N_{\nu,m}$, the total sum of the massive neutrinos $\Sigma m_{\nu}$ and the dark energy equation of state parameter $w_\mathrm{de}$ (corresponding in \class\ to $\Omega_{cdm}= \Omega_m - \Omega_b$ since $\Omega_b$ is fixed, $N_{ur}$, $N_{ncdm}$, $m_{ncdm}$ = $\Sigma m_{\nu} / N_{\nu,m}$ and $w_{0,fld}$) change in different models and are reported in the corresponding sections. $\Omega_\Lambda$ is computed by imposing the flatness condition $\sum_i \Omega_i = 1$.

\section{A simple model: Dark Energy with a constant equation of state}
\label{sec:wcdm}

\subsection{Equations}
In order to search for deviations from a cosmological constant, we consider here a dark energy equation of state with $w_\mathrm{de}=p_\mathrm{de}/\rho_\mathrm{de}\neq -1$. The equations for this model have been studied in detail in~\cite{Ballesteros:2010ks}, and experimental constraints have been presented in, e.g.,~\cite{Tripathi:2016slv,Scolnic:2017caz,Aghanim:2018eyx,Abbott:2021bzy}. 

For a constant dark energy equation of state $w_\mathrm{de}$, the dark energy density is given by\footnote{In the code, $w_\mathrm{de}$ is written as the parameter \texttt{w0}.}
\begin{equation}
    \rho(a)=\rho_0\cdot a^{-3(1+w_\mathrm{de})}.
\end{equation}
We immediately see that $\rho=\textrm{const}$. for a cosmological constant with $w_\mathrm{de}=-1$. The Hubble parameter is given by
\begin{equation}
    \frac{H(a)}{H_0}=\Big[\Omega_r a^{-4}+\Omega_m a^{-3}+\Omega_\kappa a^{-2}+\Omega_\Lambda\cdot a^{-3(1+w_\mathrm{de})} \Big]^{\frac{1}{2}} . 
\end{equation}
In this equation, $\Omega_r$ is the radiation density which includes photons and massless neutrinos, $\Omega_m$ the matter density, $\Omega_\kappa$ the curvature density (listed for completeness, even though the Boltzmann solver in \PyCosmo\ currently only supports flat models) and $\Omega_\Lambda$ the dark energy density. In all cases, $\Omega_i$ is defined as the fraction of the energy density of the corresponding component today and the critical energy density $\rho_\mathrm{crit}$ of the Universe.

In the case of a cosmological constant, there are no dark energy perturbations. For $w_\mathrm{de}\neq -1$, we can write down the dark energy equations following 
\cite{ma,dodelson} and obtain
\begin{align}
\begin{split}
    \label{eq:perturbations1}
    \dot{\delta}_\mathrm{de}&=-(1+w_\mathrm{de})(ku_\mathrm{de}+3\dot{\Phi})-3\frac{\dot{a}}{a}\Big(\frac{\delta p_\mathrm{de}}{\delta\rho_\mathrm{de}}-w_\mathrm{de}\Big)\delta_\mathrm{de}\\
    \dot{u}_\mathrm{de}&=-\frac{\dot{a}}{a}\Big(1-3w_\mathrm{de}\Big)u_\mathrm{de}+\frac{\delta p_\mathrm{de}/\delta\rho_\mathrm{de}}{1+w_\mathrm{de}}k\delta_\mathrm{de}-k\sigma_\mathrm{de}+k\Psi,
\end{split}
\end{align}
where all derivatives are with respect to the conformal time $\eta$ and we use the conformal Newtonian gauge as in \cite{pycosmo}. The anisotropic stress, $\sigma_\mathrm{de}$, vanishes, which deletes one term in the second perturbation equation.  

In general, the sound speed $\tilde{c}_\mathrm{s,de}^2=\frac{\delta p_\mathrm{de}}{\delta\rho_\mathrm{de}}$ is a Gauge-dependent variable, which can be expressed in terms of the rest frame sound speed $c_\mathrm{s,de}^2$ and the adiabatic sound speed $c_\mathrm{a}^2$. The latter is equal to $w_\mathrm{de}$ for a constant dark energy equation of state \citep{Ballesteros:2010ks}. Here, we use the expression \citep{Ballesteros:2010ks}
\begin{equation}
    \tilde{c}_\mathrm{s,de}^2\delta_\mathrm{de}=c_\mathrm{s,de}^2\delta_\mathrm{de}+3\frac{d\ln{a}}{d\eta}(1+w_\mathrm{de})(c_\mathrm{s,de}^2-w_\mathrm{de})\frac{u_\mathrm{de}}{k},
\end{equation}
which is valid for a constant dark energy equation of state.

Inserting this expression in Eq.~\ref{eq:perturbations1} we obtain 
\begin{align}
\begin{split}
    \label{eq:de_perturb}
    \dot{\delta}_\mathrm{de}&=-(1+w_\mathrm{de})(ku_\mathrm{de}+3\dot{\Phi})-3\frac{\dot{a}}{a}\Big(c_\mathrm{s,de}^2-w_\mathrm{de}\Big)\delta_\mathrm{de}-9\Big(\frac{\dot{a}}{a}\Big)^2(1+w_\mathrm{de})(c_\mathrm{s,de}^2-w_\mathrm{de})\frac{u_\mathrm{de}}{k}\\
    \dot{u}_\mathrm{de}&=-\frac{\dot{a}}{a}(1-3c_\mathrm{s,de}^2)u_\mathrm{de}+\frac{c_\mathrm{s,de}^2}{1+w_\mathrm{de}}k\delta_\mathrm{de}+k\Psi. 
\end{split}
\end{align}

Compared to the system of equations in \cite{pycosmo},
the Einstein equations are modified to 
\begin{eqnarray}
k^{2}\Phi+3 \frac{\dot{a}}{a} \left( \dot{\Phi}- \frac{\dot{a}}{a} \Psi \right) &  = &  4 \pi G a^2 \left[ \rho_m \delta_m + 4 \rho_r \Theta_{r0} + \rho_\mathrm{de} \delta_\mathrm{de}\right] \label{eq:eb_phi_tt}\\
\dot{\Phi}- \frac{\dot{a}}{a} \Psi & = & - 4\pi G \frac{a^2}{k} \left[ \rho_m u_m + 4 \rho_r \Theta_{r1} + \rho_\mathrm{de}(1+w_{\mathrm{de}}) u_\mathrm{de} \right]\\
k^2 (\Phi+\Psi) & = & -32 \pi G a^2 \rho_r \Theta_{r2} . 
\end{eqnarray}

In order to evolve the perturbation equations, we also need to define the initial conditions for these. We choose the adiabatic initial conditions from \class, outlined in~\cite{Ballesteros:2010ks} in the synchronous gauge, and then transform them into the conformal Newtonian gauge. We report the initial conditions for all the fields in Appendix~\ref{app:ini}.

\subsection{Numerical implementation}
In order to implement the new equations outlined in the previous section, we generated a new file for the symbolic Boltzmann equations, called \texttt{CosmologyCore\_wcdm.py}. 
This can be used instead of the default equations file \texttt{CosmologyCore.py} for the \LCDM model, by setting the \texttt{model} to \texttt{"wcdm"} as shown in Listing \ref{ls:cosmo}.

In \PyCosmo, all derivatives are written with respect to $\ln{a}$ (in the code \texttt{lna}) instead of $\eta$. The perturbation equations for a $\Lambda$CDM model in this notation are detailed in~\ref{app:eqns}. 
In the previous section, we presented the dark energy perturbation equations with respect to $\eta$. Using the conversion  $\frac{d\ln{a}}{d\eta}= aH$, we can rewrite the two dark energy perturbation equations from Eq.~\ref{eq:de_perturb} as
\begin{align}
\begin{split}
    \label{eq:de_perturb_lna}
    \frac{d\delta_\mathrm{de}}{d\ln{a}}&=-(1+w_\mathrm{de})\Big(\frac{ku_\mathrm{de}}{aH(a)}+3\frac{d\Phi}{d\ln{a}}\Big)-3(c_\mathrm{s,de}^2-w_\mathrm{de})\Big(\delta_\mathrm{de}+3(1+w_\mathrm{de})\frac{aH(a)u_\mathrm{de}}{k}\Big)\\
    \frac{du_\mathrm{de}}{d\ln{a}}&=-(1-3c_\mathrm{s,de}^2)u_\mathrm{de}+\frac{c_\mathrm{s,de}^2}{1+w_\mathrm{de}}\frac{k\delta_\mathrm{de}}{aH(a)}+\frac{k}{aH(a)}\Psi. 
\end{split}
\end{align}

Then the two dark energy perturbation equations in Eq.~\ref{eq:de_perturb_lna} can be expressed with \sympy as

\begin{lstlisting}[caption={The two dark energy perturbation equations as implemented in \texttt{CosmologyCore\_wcdm.py}.},captionpos=b, label=ls:de_perturb_sympy]
ddelta_de_dlan =  - (1 + w0) * (k / (a * H) * u_de + 3 * dPhi_dlan) 
        - 3 * (cs_de2 - w0) * delta_de
        - 9 * (1 + w0) * (cs_de2 - w0) * (a * H) / k * u_de 
    
du_de_dlan =  - (1 - 3 * cs_de2) * u_de
        + cs_de2 / (1 + w0) * k * delta_de / (a * H)
        + k / (a * H) * Psi 
\end{lstlisting}
\texttt{CosmologyCore\_wcdm.py} also contains the background equations and the initial conditions for the linear perturbations. 
\subsection{Code comparisons}

\begin{figure}
    \centering
    \includegraphics[scale=0.45]{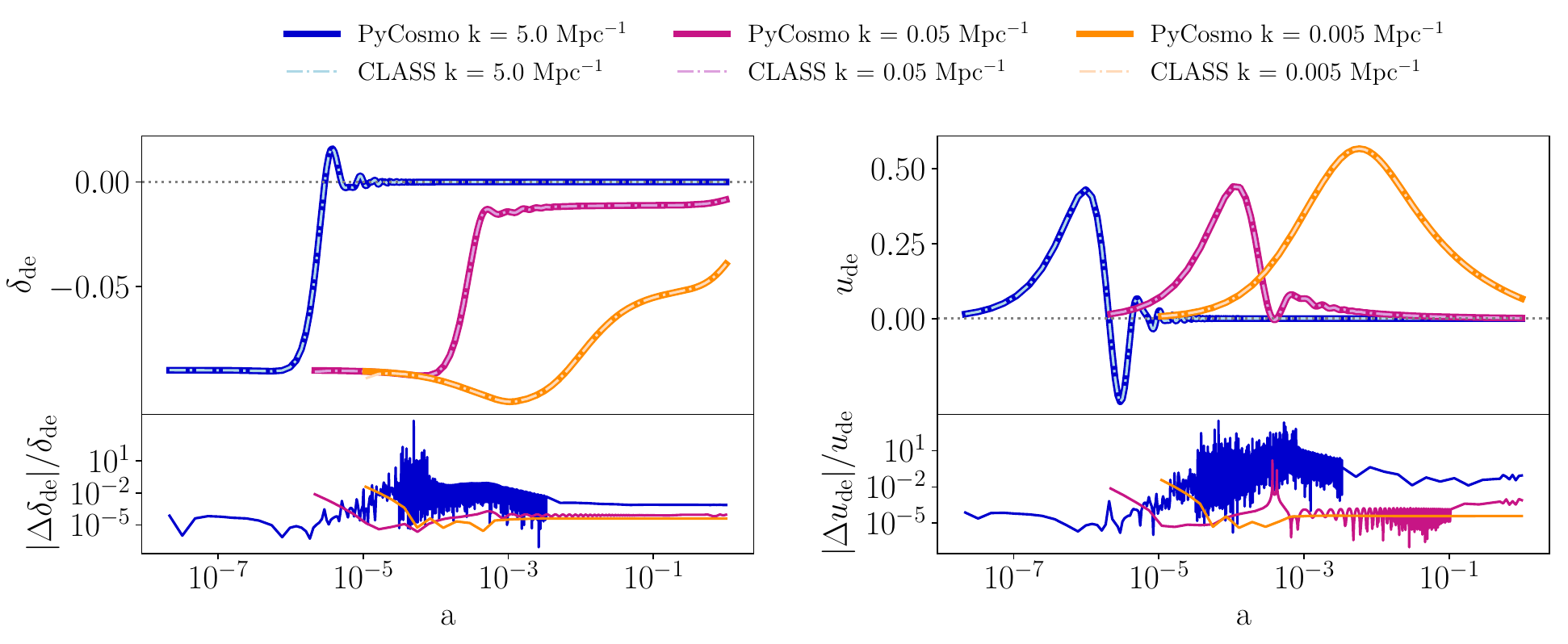}
    \caption{In the top left panel, we show the dark energy perturbation $\delta_\mathrm{de}$ as a function of scale factor $a$ for a dark energy equation of state with $w=-0.9$ and sound speed $c_\mathrm{s,de}^2=1$ at three values of the wavenumber $k=0.005,0.05$ and $5\ \mathrm{Mpc}^{-1}$. In the top right panel, we show dark energy perturbation $u_\mathrm{de}$ as a function of scale factor $a$ for the same wavenumbers. Perturbations computed with \PyCosmo\ are displayed by thick lines, while \class\ values by dash-dotted lines. The bottom panels display relative differences between the two codes.}
    \label{fig:wcdm_fields}
\end{figure}

\begin{figure}
    \centering
    \includegraphics[width=0.8\linewidth]{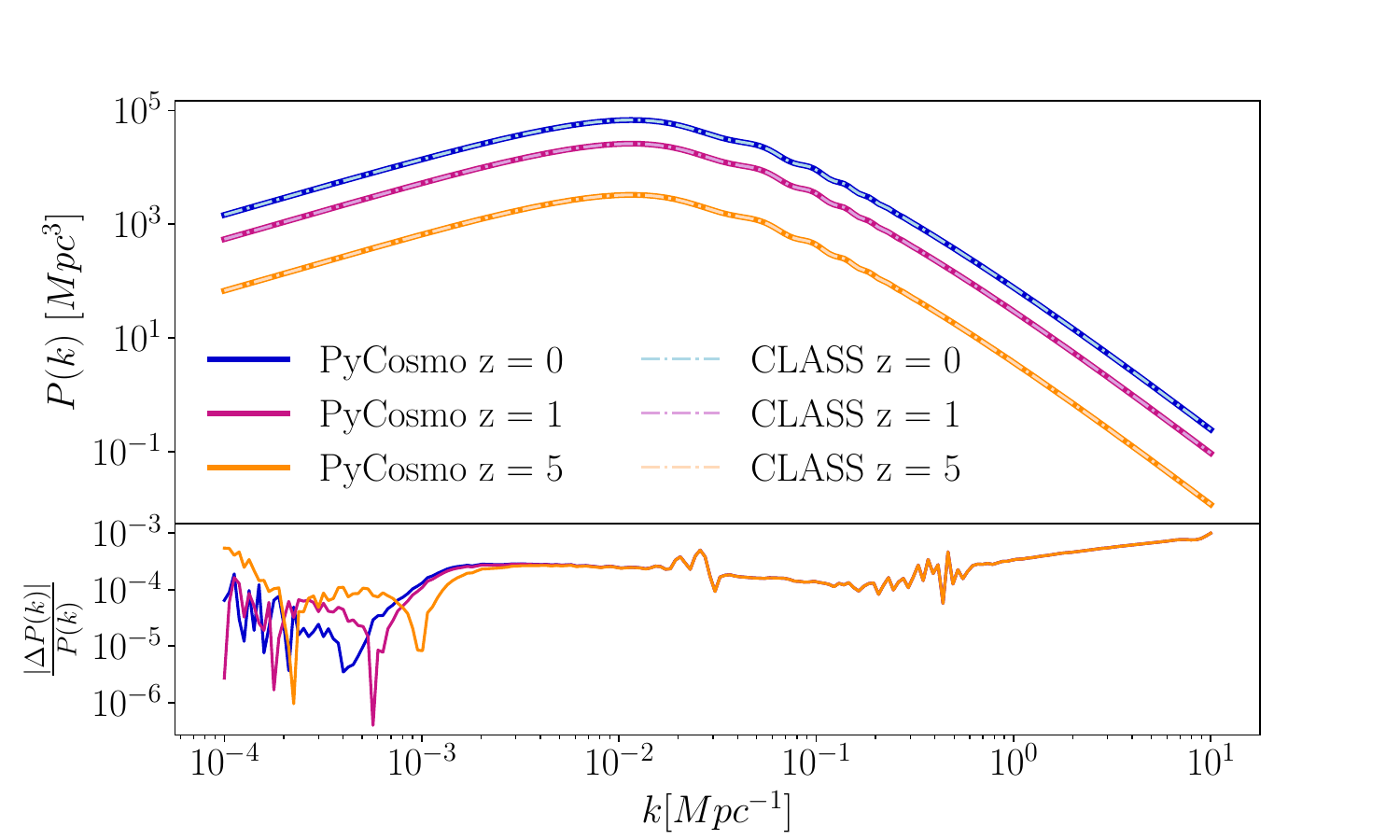}
    \caption{Total matter power spectrum at redshifts $z=0$, $z=1$ and $z=5$ for a dark energy equation of state with $w=-0.9$ and sound speed $c_\mathrm{s,de}^2=1$, plotted both with \PyCosmo\ (thick solid lines) and \class\ (dash-dotted lines) for 200 log-spaced $k$ values between $10^{-4}$ and $10\ \mathrm{Mpc^{-1}}$. On the bottom panel we display the relative difference between the two codes.}
    \label{fig:wcdm_pk}
\end{figure}

In Figure~\ref{fig:wcdm_fields}  we show the dark energy perturbations $\delta_\mathrm{de}$ and $u_\mathrm{de}$ as a function of scale factor $a$ for three wavenumbers $k=0.005,0.05$ and $5\ \mathrm{Mpc}^{-1}$, plotted both with \PyCosmo\ as well as \class. We also display the relative difference between the evolution of the perturbations obtained by the two codes. The cosmology we consider is 
\begin{itemize}
    \item \textbf{$w$CDM}:  \{$\Omega_\Lambda$, $\Omega_m$, $N_{\nu}$, $N_{\nu,m}$, $\Sigma m_{\nu}$, $w_\mathrm{de}$\} = \{0.69992, 0.3, 3.044, 0, 0, -0.9\}
\end{itemize} 
with all other parameters as specified in \ref{app:constants} and using the \textit{precision} settings from \ref{sec:precision}. In general, we find good agreement between the codes. When the fields are highly oscillating around zero, we observe a degradation of the agreement, as expected, given the impact of step-size control and numerical precision of the solver in that regime. We also notice a discrepancy at initial time for small values of $k$, which is caused by the tight coupling approximation in \class. In Figure~\ref{fig:wcdm_pk} we show the wCDM total matter power spectrum computed with the two codes for 200 log-spaced $k$ values between $10^{-4}$ and $10\ \mathrm{Mpc^{-1}}$ at redshifts $z=0$, $z=1$ and $z=5$. In general, we observe that the results for $w$CDM show the same level of agreement with \class\ as $\Lambda$CDM. The discrepancies tend to grow on large scales for $z=5$ but remain below the $10^{-3}$ level. The same is observed for higher redshifts.
In Section \ref{sec:discuss} we summarize the comparisons in terms of computing time and power spectrum relative difference for different $k$ ranges at $z=0$. 

\section{A complex model: Massive Neutrinos}
\label{sec:massive_nu}

\subsection{Equations}
Oscillation experiments provide evidence that neutrinos have mass (see e.g.~\cite{PhysRevLett.81.1562,PhysRevLett.89.011301,PhysRevLett.90.021802} and also \cite{de_Salas_2021} for a recent global fit of neutrino oscillation data) and since their masses are imprinted onto cosmological observables, we need to include the evolution of light massive relics in the system of equations. This allows cosmological probes to constrain the properties of neutrinos, in particular the sum of the neutrino masses (see e.g.  \cite{Lesgourgues:2006nd,lesgourgues_mangano_miele_pastor_2013,Lattanzi:2017ubx} for reviews on neutrino cosmology). 

In this section, we present the implementation of the Einstein-Boltzmann equations for massive neutrinos into the \PyCosmo\ Boltzmann solver. 
Massive neutrinos modify both the background evolution and the linear order perturbations. Qualitatively, massive neutrinos undergo a phase transition: they behave like radiation at early times, when they are fully relativistic, and shift to a matter-like behaviour at late times (see, e.g., \cite{Lattanzi:2017ubx,Ichikawa:2004zi}). The transition happens smoothly through cosmic time and the dependence on mass, scale factor and momentum in the evolution of the distribution function prevents from integrating out the momentum dependence. This can be done only when considering approximations. For this reason, the inclusion of massive neutrinos in the Boltzmann equations is highly non trivial and has a strong impact on the size of the ODE system.

In this section we write the equations for $N_{\nu,m}$ massive neutrinos with degenerate masses and total neutrino mass sum $\Sigma m_\nu$, where the sum goes over the three neutrino mass eigenstates. A generalization to neutrinos with different masses is simply achieved by suppressing the $N_{\nu,m}$ in front of the equations and writing separate equations for each neutrino species with mass $m_{\nu,i}$. In \PyCosmo\ we introduce degenerate massive neutrinos. The introduction of neutrino hierarchies is left as future development.

The massive neutrino density can be written as 
\begin{equation}
\rho_{\nu,m}(a) = 4 \pi a^{-4} \int_0^\infty dq q^2 f_0(q) \epsilon,
\label{eq:rho_nu_m}
\end{equation} 
where $q = ap$, with $p$ the proper momentum, related to the 4-momentum by $P_i = a (1-\Phi)p_i$ in the Newtonian conformal gauge. Then $\epsilon = \sqrt{q^2+a^2 m^2}$ is the proper energy measured by a comoving observer multiplied by the scale factor and $f_0(q)$ is the Fermi-Dirac distribution 
\begin{equation}
f_0(q)=\frac{g_s}{(2\pi\hbar)^3}\frac{1}{e^{q/k_bT_0}+1},
\label{eq:fd}
\end{equation}
where $g_s$ is the spin degeneracy factor that equals $2 N_{\nu,m}$ in the case of degenerate neutrinos. $T_0$ is the temperature of the Cosmic Neutrino Background today expressed in units of the CMB temperature $T_0 = T_{{\nu},m} T_{\mathrm{CMB}}$, $T_{{\nu},m} = \left(\frac{4}{11}\right)^{1/3}$ for neutrinos that undergo instantaneous decoupling and $k_b$ is the Boltzmann constant.
The Friedmann equation results in
\begin{equation}
\frac{H(a)}{H_{0}} =  \left[   (\Omega_{r}+ \Omega_{\nu,m}(a) ) a^{-4} + \Omega_{m} a^{-3} + \Omega_{\kappa} a^{-2} 
+ \Omega_{\Lambda} \right]^{\frac{1}{2}}
\end{equation} 
where $\Omega_r$ still includes massless neutrinos if present (denoted with the subscript $\nu$ such that $\Omega_r = \Omega_\gamma + \Omega_\nu$), whereas the massive neutrino energy density is $\Omega_{\nu,m}(a)$.
Note that we factor out the $a^{-4}$ term from $\Omega_{\nu,m}(a)$ for similarity with the other energy densities, but $\Omega_{\nu,m}(a)$ still contains a dependency on the scale factor $a$, differently from the other species, since the proper energy $\epsilon$ contains a factor of $a$ that we cannot integrate out (see equation \ref{eq:rho_nu_m}, $\Omega_{\nu,m}(a) a^{-4} = \rho_{\nu,m}(a) /\rho_{crit}$).

The massive neutrino perturbations arise from a linear expansion of the distribution function $f(x,q,\hat{q},\eta) = f_0(q)\left[1+\mathcal{M}(x,q,\hat{q},\eta)\right]$ around the Fermi-Dirac distribution $f_0(q)$. The function $\mathcal{M}(x,q,\hat{q},\eta)$ is Fourier transformed and expanded in a Legendre series as 
\begin{equation}
 \mathcal{M}(\vec{k},\hat{q},q,\eta) = \sum_{l=0}^{\infty} (-i)^l (2l+1)\mathcal{M}_l(\vec{k},q,\eta)P_l(\mu)
 \label{eq:legendre}
 \end{equation} 
with $\mu = \hat{k} \cdot \hat{q}$, $\mathcal{P}_l$ the Legendre polynomials and $\mathcal{M}_l$ defined as 
\begin{equation}
\mathcal{M}_l=\frac{1}{(-i)^l} \int_{-1}^{1} \frac{d\mu}{2} {\mathcal P}_l(\mu)\mathcal{M}(\mu).
\end{equation}  
The massive neutrino Boltzmann equations are then derived similarly to those of the ultra-relativistic fields, setting the collision term to $0$, since neutrinos are only weakly interacting. Using the definition of $\mathcal{M}_l$ to express the Boltzmann equations as a hierarchy of moments, we obtain
\begin{align}
\dot{\mathcal{M}}_0 &= -\frac{qk}{\epsilon}\mathcal{M}_1-\dot{\Phi}\frac{d\ln{f_0}}{d\ln q}\\
\dot{\mathcal{M}}_1 &= \frac{qk}{3\epsilon}\left(\mathcal{M}_0-2\mathcal{M}_2\right)-\ \frac{ \epsilon k}{3q} \Psi \frac{d\ln{f_0}}{d\ln{q}}\\
\dot{\mathcal{M}}_l &= \frac{qk}{(2l+1)\epsilon}\left[l\mathcal{M}_{l-1}-(l+1)\mathcal{M}_{l+1}\right],\	\	l\geq 2.
\end{align}
The hierarchy is truncated at a multipole $l_{max}$ (\texttt{l\_max\_mnu} in the code) when solving the system of equations numerically with a hierarchy truncation from \cite{ma}, which is analogous to that for photons and massless neutrinos
\begin{equation}
\dot{\mathcal{M}}_{l_{max}} \simeq \frac{qk}{\epsilon} \mathcal{M}_{l_{max}-1} - \frac{(l_{max}+1)}{\eta}\mathcal{M}_{l_{max}}.
\end{equation}
Massive neutrinos also modify the Einstein equations due to the extra terms in the stress-energy tensor. These now read
\begin{small}
\begin{align}\label{eq:massive_einstein}
k^2(\Psi + \Phi) &= - 12 \left(\frac{H_0}{a}\right)^2\left(\Omega_{\gamma}\Theta_2 + \Omega_{\nu}\mathcal{N}_2 + \frac{N_{\nu,m}}{(2\pi)^2 \rho_{crit} }\int{q^2 dq \frac{q^2}{\epsilon}\frac{\mathcal{M}_2}{e^{q/k_b T_0}+1}}\right)\\
\nonumber k^2 \Phi + 3\frac{\dot{a}}{a}\left(\dot{\Phi}-\frac{\dot{a}}{a}\Psi\right) &= \frac{3}{2}\left(H_0a\right)^2\left[\Omega_m \delta_m  a^{-3} + 4 a^{-4}  \left(\Omega_{r} \Theta_{r0} + \frac{N_{\nu,m}}{(2\pi)^2\rho_{crit}}\int{q^2 dq \frac{\epsilon \mathcal{M}_0}{e^{q/k_b T_0}+1}}\right)\right],
\end{align}
\end{small}
\noindent where $\Omega_m\delta_m$ is a shortcut for $\Omega_{dm}\delta + \Omega_b \delta_b$, $\Omega_r \Theta_{r0} = \Omega_{\gamma}\Theta_0 + \Omega_{\nu}\mathcal{N}_0$ and $\rho_{crit}$ is the critical energy density of the Universe.
The massive neutrino quantities we are interested in are the density fluctuation, the fluid velocity and the shear stress which are computed by integrating over the moments $\mathcal{M}_l$:  
\begin{align}\label{eq:tmunu_massivenu}
\delta_{\nu,m} &= \frac{\int{q^2 dq \epsilon f_0(q) \mathcal{M}_0}}{\int{q^2 dq \epsilon f_0(q)}}\\
\nonumber u_{\nu,m} &= \frac{\int{q^2 dq q f_0(q)\mathcal{M}_1}}{\int{q^2 dq \epsilon f_0(q)}+\frac{1}{3}\int{q^2 dq \frac{q^2}{\epsilon}f_0(q)}}\\
\nonumber\sigma_{\nu,m} &= \frac{2}{3}\frac{ \int{q^2 dq \frac{q^2}{\epsilon} f_0(q)\mathcal{M}_2}}{\int{q^2 dq \epsilon f_0(q)}+\frac{1}{3}\int{q^2 dq \frac{q^2}{\epsilon}f_0(q)}}.
\end{align}
Note that in the \PyCosmo\ implementation, we substitute $q$ with $q' = \frac{q}{k_b T_0}$ for convenience.

\subsection{Numerical implementation}

The implementation of the massive neutrino equations uses \sympytoc similarly to the \PyCosmo\ implementation of \LCDM and $w$CDM. The background integral over momentum $q$ is computed using indefinite numerical integration from \sympytoc, whereas the integrals at perturbation level use a Gauss-Laguerre quadrature integration scheme in our symbolic representation of the ODE system, since we need to evolve a finite number of equations.
This follows  the approach used in \cite{Lesgourgues_2011}. The number of discrete $q$ values is governed by the parameter \texttt{mnu\_relerr}, which sets the relative difference between the Gauss-Laguerre integration of a test function ($\sum_{n=2}^{n=4} q^n f_0(q)$) with respect to its analytical result. This parameter is specified using \texttt{PyCosmo.build} since it affects the size of the ODE system and thus also code generation. 
We introduce an additional parameter influencing code generation in addition to the truncation parameter \texttt{l\_max} of the photon and massless neutrino hierarchies: \texttt{l\_max\_mnu} that truncates the Legendre series  $\mathcal{M}_l$ as described above.

The implementation of massive neutrino cosmologies results in a large systems of equations and thus \C functions with millions of lines of generated code, challenging the optimizer of the used \C compiler. To mitigate significantly compilation time and memory requirements, we enable \sympytoc to use the matrix splitting feature (enabled by specifying the \texttt{splits} parameter and \texttt{reorder=True} in \texttt{PyCosmo.build}) described in Section \ref{sec:pycosmo2}. The user can also pass a \texttt{compilation\_flags} parameter, that enables or disables compiler optimizations of the \C code and has a diametrical effect on compilation vs. runtime.

We implement initial conditions for massive neutrinos that match the adiabatic initial conditions in \class\ and \texttt{Cosmics} and can be triggered using the \texttt{initial\_conditions} parameter. We report the equations in Appendix~\ref{app:ini}.

\subsection{Code comparisons}
\label{sec:massive_nu_cc}
\begin{figure}[h]
    \centering
    \includegraphics[width=0.8\linewidth]{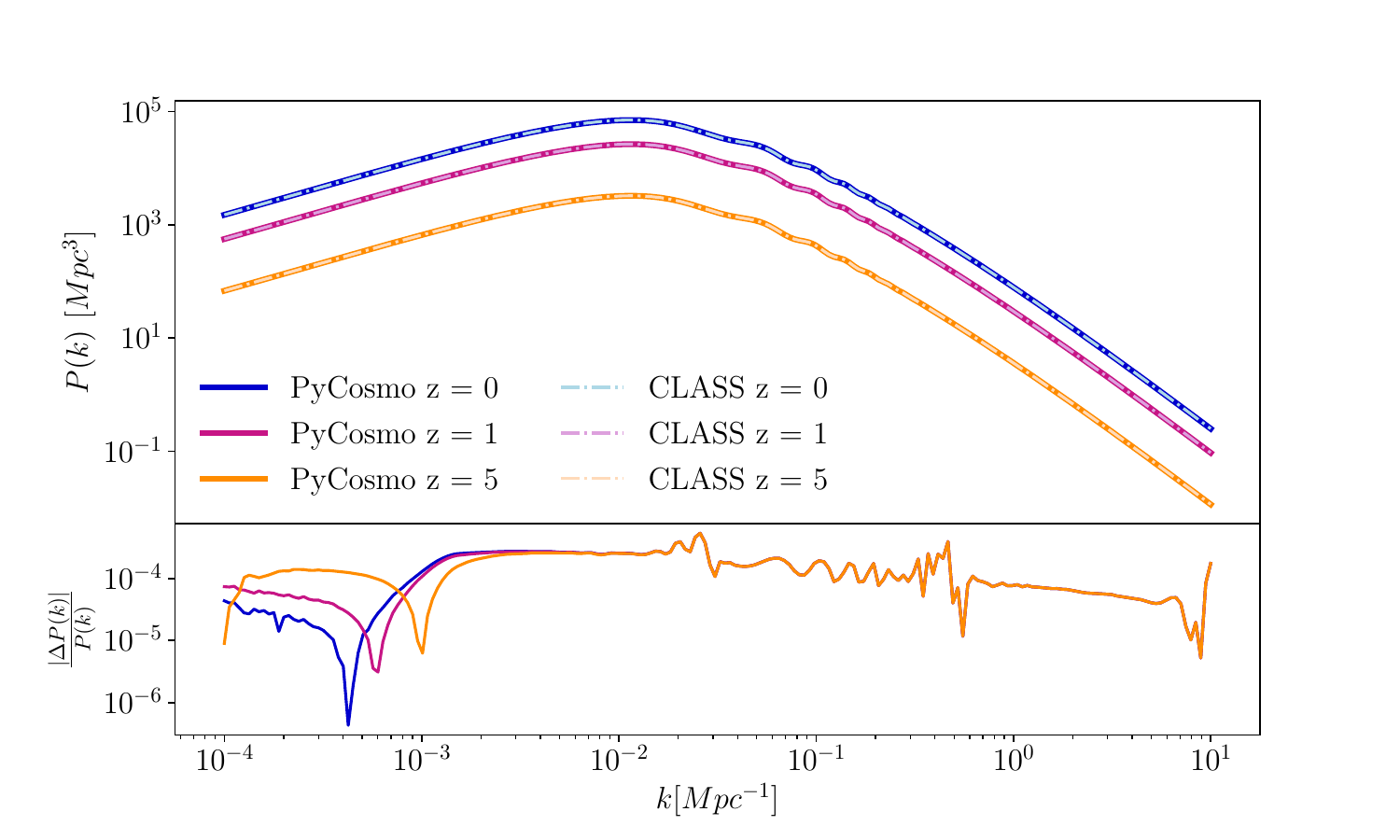}
    \caption{Total matter power spectrum at redshifts $z=0$, $z=1$ and $z=5$ for a cosmology with three massive neutrinos with $\Sigma m_{\nu}$ = 60 meV, plotted both with \PyCosmo\ and \class\ for 200 log-spaced $k$ values between $10^{-4}$ and $10\ \mathrm{Mpc^{-1}}$. In the bottom panel, we display the relative difference between the two codes.}
    \label{fig:mnu_pk}
\end{figure}

\begin{figure}[h]
    \centering
    \includegraphics[width=\linewidth]{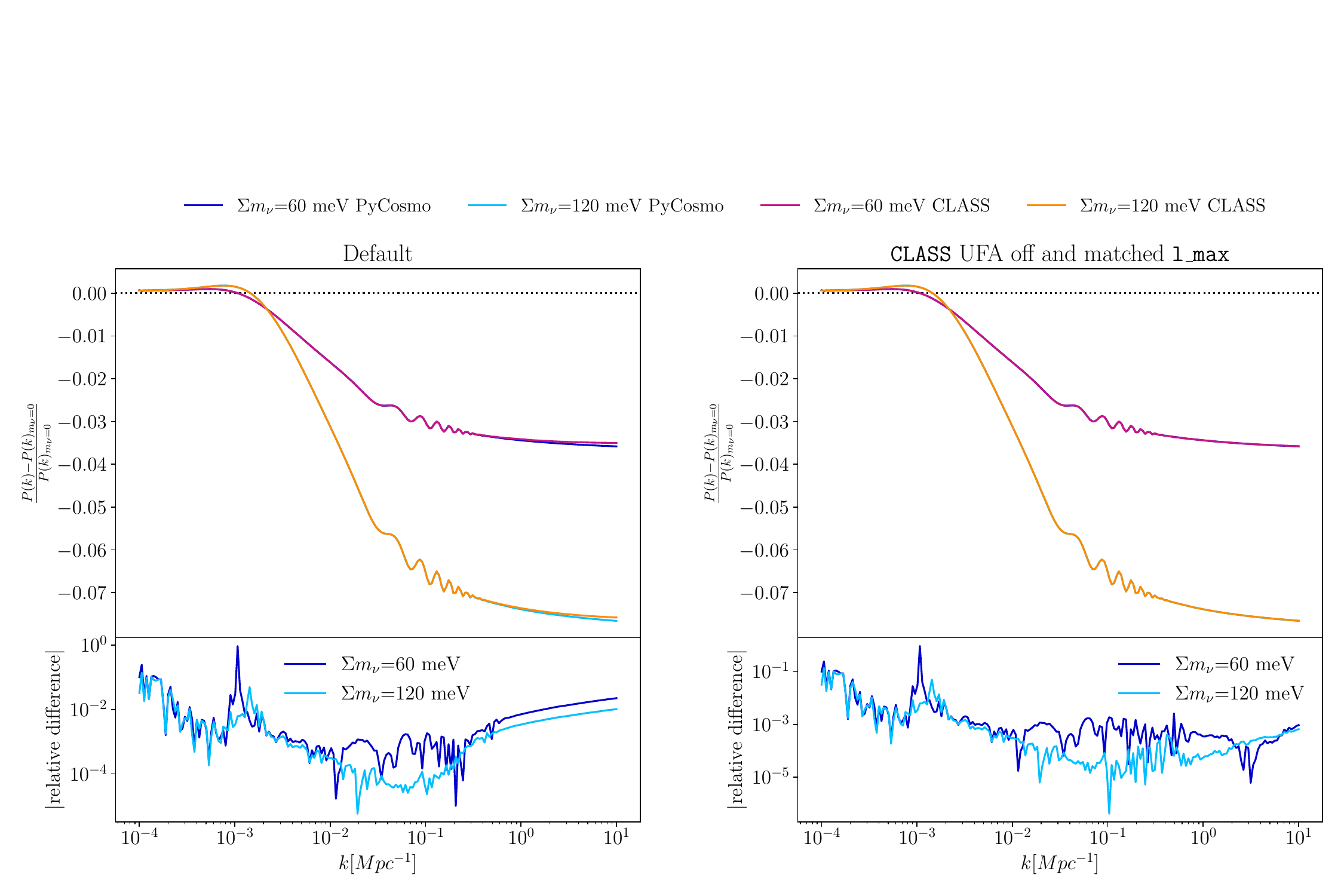}
    \caption{Suppression of the total matter power spectrum for two cosmologies with three neutrinos and $\Sigma m_{\nu}$ = 60 and 120 meV, respectively, compared to a $\Lambda$CDM cosmology with massless neutrinos only computed with \texttt{PyCosmo} and \class. For all cosmologies,  we set the total matter density to $\Omega_{m,tot}$ = 0.3 and $N_{\mathrm{eff}}=3.044$. The left panel displays the power spectrum and the relative difference with the default \textit{precision} settings for \class\ and \PyCosmo. In the right plot, we display the same quantities when setting \texttt{l\_max = 50} for all the radiation fields and massive neutrinos and switching off the UFA in \class.}
    \label{fig:mnu_suppress}
\end{figure}

One of the key effects of massive neutrinos on cosmological observables is the suppression of small scale matter overdensities due to neutrino free streaming \citep{Lesgourgues:2006nd}. In Figure \ref{fig:mnu_pk} we show the total matter power spectrum obtained with \PyCosmo\ and \class\ for 200 log-spaced $k$ values between $10^{-4}$ and $10\ \mathrm{Mpc^{-1}}$ at redshifts $z=0$, $z=1$ and $z=5$ with the following cosmological parameters:
\begin{itemize}
    \item \textbf{degenerate $\Sigma m_\nu$ = 60 meV}:  \{$\Omega_m$, $N_{\nu}$, $N_{\nu,m}$, $\Sigma m_{\nu}$, $w_\mathrm{de}$\} = \{0.29869, 0.00440, 3, 0.06, -1\}\footnote{$\Omega_m$ and $N_{\nu}$ are determined by fixing $\Omega_{m,tot}=\Omega_m + \Omega_{\nu,m}=0.3$ and $N_\mathrm{eff}=N_{\nu}+{T_{\nu,m}}^4\left(\frac{4}{11}\right)^{-\frac{4}{3}}N_{\nu,m}\ =$ 3.044 in order to look at the effects of neutrino mass on the total matter power spectrum in Figure \ref{fig:mnu_suppress}.}.
\end{itemize}
All other parameters are set to default values (see \ref{app:constants}) and we use the \textit{precision} settings for the two codes (see \ref{sec:precision}).
In the bottom panel we display the  relative difference between \PyCosmo\ and \class.
In Appendix~\ref{app:pk} we outline the equation of the total matter power spectrum, following a fully general relativistic treatment in the presence of massive neutrinos \citep{Yoo_2009, Yoo_2010, Bonvin_2011, Challinor_2011,Dio_2013}. 
In general, we observe a very good agreement, with a maximum relative discrepancy of $5\times 10^{-4}$ on intermediate scales. The redshift evolution does not impact the agreement.  We also show the suppression of the total matter power spectrum with respect to the power spectrum with massless neutrinos in Figure \ref{fig:mnu_suppress} both for $\Sigma m_\nu$ = 60 meV and 120 meV ($\Omega_m$=0.29737), when keeping $\Omega_{m,tot}= \Omega_m + \Omega_{\nu,m} $ fixed to $\Omega_{m,tot}=0.3$ and $N_\mathrm{eff}=3.044$.  The cosmological parameters for the $\Lambda$CDM model are set to:
\begin{itemize}
    \item \textbf{$\Lambda$CDM}:  \{$\Omega_m$, $N_{\nu}$, $N_{\nu,m}$, $\Sigma m_{\nu}$, $w_\mathrm{de}$\} = \{0.3, 3.044, 0, 0., -1\}.
\end{itemize}
On the left panel of the figure, we show the suppression of the power spectrum computed with \PyCosmo\ and \class, using the default \textit{precision} settings and the relative difference between the two codes. The suppression has a large discrepancy at low $k$ values when it is approaching and crossing zero. Furthermore, there is a $\sim1\%$ difference on small scales (large $k$ values), where the effects of the hierarchy truncation and the approximations are most dominant. We verify that this discrepancy is reduced when matching the \texttt{l\_max} parameters in the two codes (\texttt{l\_max\_g = l\_max\_pol\_g = l\_max\_ur = l\_max\_ncdm = 50} in \class, \PyCosmo\ remains in \textit{precision} settings) and suppressing the Ultra-relativistic Fluid Approximation in \class, as shown in the right panel of Figure~\ref{fig:mnu_suppress}.

\section{An approximation scheme: Radiation Streaming Approximation}
\label{sec:RSA}
\subsection{Equations}
After decoupling, photons and massless neutrinos behave approximately like test particles free-streaming in the gravitational field determined by the massive components, making it possible to derive a non-oscillatory solution of the inhomogeneous Boltzmann equations inside the Hubble radius. This approximation, called the Radiation Streaming Approximation (RSA), was introduced in the Newtonian gauge by \cite{Doran:2005ep} and in the synchronous gauge by \cite{Blas_2011}. This treatment allows both to avoid unphysical oscillations resulting from the hierarchy truncation and to speed up the integration, which is slowed down by fast late time oscillations of the radiation fields especially on small scales. At late times, the approximation does not need to be precise, since the contribution of the radiation energy density to the overall energy density is negligible.
This approximation only impacts the linear perturbations. The evolution of the relativistic fields can be written as
\begin{align}
\begin{split}
\Theta_0 &= \Phi + \frac{\dot{\tau}u_b}{k}\\
\Theta_1 &= -\frac{2}{k}\dot{\Phi} + \frac{\dot{\tau}}{k}\left( \frac{aH}{k}u_b - c_b^2 \delta_b + \Phi\right)\\
\mathcal{N}_0 &= \Phi \\
\mathcal{N}_1 &= - \frac{2}{k}\dot{\Phi}\\
\Theta_2 &= \Theta_{P0,1,2} = \mathcal{N}_2 = 0 ,\\
\end{split}
\end{align}
with all the higher order multipoles set to 0.
This approximation is switched on when two conditions are satisfied, following the \class\ \citep{Blas_2011} scheme: 
\begin{itemize}
\item $  \texttt{rsa\_trigger\_k\_eta} \leq k \eta$
\item $  \texttt{rsa\_trigger\_taudot\_eta} \leq -(\dot{\tau}\eta)^{-1}$,
\end{itemize}
corresponding to decoupled radiation within the horizon.

\subsection{Numerical implementation}
In order to implement the Radiation Streaming Approximation within \PyCosmo\, we use a functionality from \sympytoc\ to switch between two different ODEs at a dynamically computed time point.

This requires:
\begin{itemize}
    \item Two ODE systems specified as  \sympytoc \texttt{OdeFast} objects. In our case these will be the symbolic representations of the model of interest ($\Lambda$CDM, $\Lambda$CDM with massive neutrinos or $w$CDM) and its equivalent RSA system. Note that the two systems can have different dimensions.
    \item A \texttt{switch\_time} function which determines at which time to switch from the first system of equations to the second. In the RSA implementation we use the switching conditions described above.
    \item A \texttt{switch} function, computing the initial conditions for the second system of equations from the state of the first system before and at the switching time. In the RSA implementation, this function just discards the matrix entries for the fields $\Theta_i$, $\Theta_{Pi}$ and $\mathcal{N}_i$, $i \ge 0$ since $\Theta_0$, $\Theta_1$, $\mathcal{N}_0$ and $\mathcal{N}_1$ have analytical expressions and thus do not need initial conditions.
    \item A \texttt{merge} function which specifies how to combine the matrix valued results from both numerical solutions into a final matrix. Our RSA implementation keeps the full matrix from the solution of the full system of equations and extends the matrix from the \texttt{RSA} solution using the analytical formula for $\Theta_i$ and $\mathcal{N}_i$ for $i \le 1$. All other entries are set to $0$ for $i \ge 2$ and for all the polarization terms $\Theta_{Pi}$.
\end{itemize}
In order to use the RSA in \PyCosmo\, one needs to pass the \texttt{rsa = True} flag to \texttt{PyCosmo.build} for any of the models. This will switch the \texttt{CosmologyCore\_}\textit{model}\texttt{.py} equation file to a \newline \texttt{CosmologyCore\_}\textit{model}\texttt{\_rsa.py}  file containing the RSA equations for the radiation fields and all the other equations of the system. 

\subsection{Internal code consistency}
\begin{figure}[h]
    \centering
    \begin{subfigure}[b]{0.8\textwidth}
    \centering
    \includegraphics[width=\linewidth]{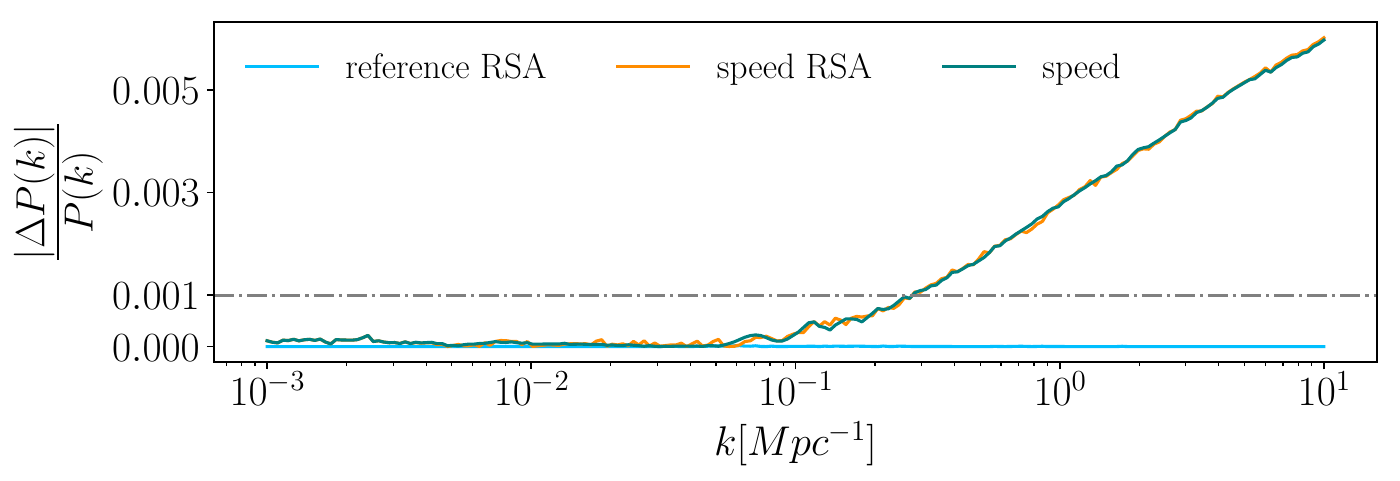}
    \caption{Relative difference between \PyCosmo\ \textit{reference} and \textit{speed} settings with (orange) and without (green) RSA. }
    \label{fig:rsa_speed}
    \end{subfigure}
    \par\bigskip
    \begin{subfigure}[b]{0.8\textwidth}
    \centering
    \includegraphics[width=\linewidth]{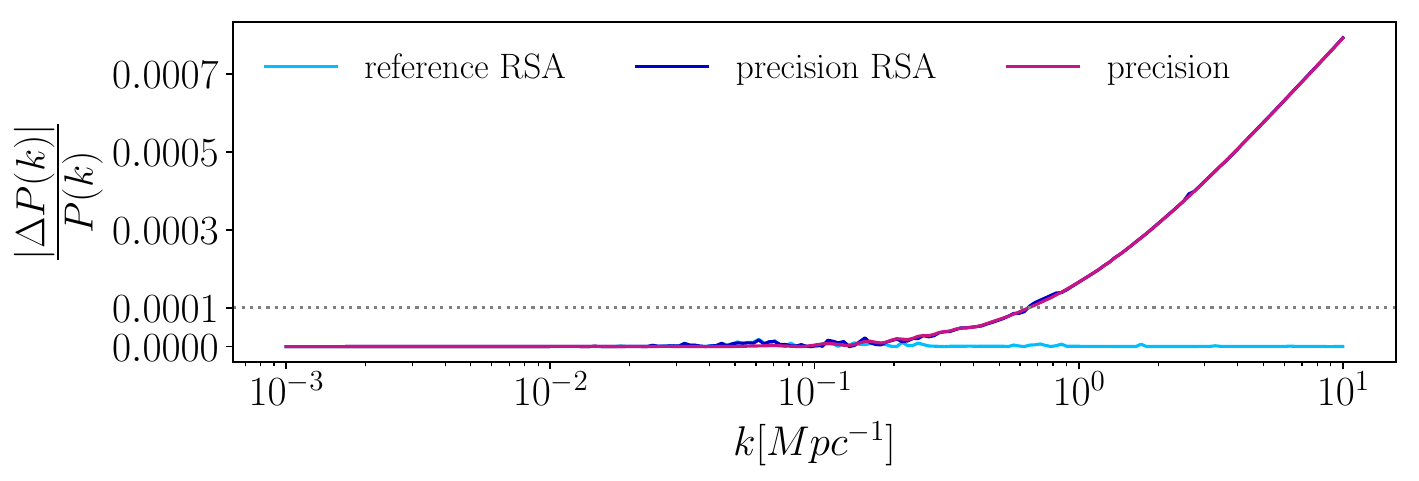}\qquad
    \caption{Relative difference between \PyCosmo\ \textit{reference} and \textit{precision}  settings with (blue) and without (purple) RSA.}\label{fig:rsa_prec}
    \end{subfigure}
\caption{Relative difference between the $\Lambda$CDM matter power spectrum computed with \texttt{l\_max} = 200 and \texttt{atol} = \texttt{rtol} = $10^{-6}$ (\textit{reference}) and the same power spectrum computed with  \PyCosmo\ \textit{speed} (top panel) and \textit{precision} (bottom panel) settings with and without RSA. We also display in both panels in light blue the relative difference between the \textit{reference} matter power spectrum with and without RSA. The grey horizontal lines correspond to a $10^{-3}$ (dash-dotted) and $10^{-4}$ (dotted)
precision level.}
\label{fig:rsa}
\end{figure}

The main achievement obtained by implementing the RSA is a significant reduction of the computation time for the fields, especially for high values of $k$, as we will discuss in detail in the next section. This is especially true when choosing the $\Lambda$CDM and $w$CDM models, where most of the system of ODEs consists of radiation perturbations.  In Figure~\ref{fig:rsa}, we show the effects of the RSA and of the hierarchy truncation on the $\Lambda$CDM matter power spectrum (\textbf{$\Lambda$CDM} cosmological parameters as specified in the Section \ref{sec:massive_nu_cc}). To show the effects of the RSA alone we display in light blue in both panels of Figure~\ref{fig:rsa} the relative difference between a \textit{reference} power spectrum with \texttt{l\_max} = 200 and \texttt{atol} = \texttt{rtol} = $10^{-6}$ and the same power spectrum computed with RSA (\texttt{rsa\_trigger\_taudot\_eta} = 100, \texttt{rsa\_trigger\_k\_eta} = 240) for 200 log-spaced $k$ values between $10^{-3}$ and $10\ \mathrm{Mpc^{-1}}$. We observe that the oscillations around zero correspond to a relative difference of the order of 0.001\%, but the computation time is reduced by approximately 80\%. In Figure~\ref{fig:rsa} we also display  the relative difference between the power spectrum computed using \PyCosmo\  \textit{speed} (panel a) and \textit{precision} (panel b) settings from Section \ref{sec:precision}, both with and without RSA, and the \textit{reference} power spectrum just described.

 Reducing the \texttt{l\_max} parameter to 50 (\textit{precision} settings) increases the discrepancy with the reference power spectrum to $\sim$0.08\% for the full system, regardless of whether the RSA is turned on or not. For the \textit{speed} settings, the discrepancy is $\sim$0.6\% both when using or not using the RSA. The computation time is again reduced by roughly 80\% when using RSA compared to solving the full equation system with the same \texttt{l\_max}, meaning that the approximation is essential to reduce the computational time of the perturbations for mid to high $k$ values without sacrificing the accuracy. The unphysical reflection of power caused by the hierarchy truncation dominates on small scales, making the inaccuracy introduced by the approximation completely negligible.

\section{Agreement and performance comparison with \class}
\label{sec:discuss}
In this section, we present benchmarks of the \PyCosmo\ Boltzmann solver for all the models described in the previous sections, both in terms of relative difference to \class\ and in terms of computing times. 
The cosmological parameters that are modified in each model are specified in the \textbf{$\Lambda$CDM}, \textbf{$w$CDM} and \textbf{degenerate $\Sigma m_\nu$ = 60 meV} parameter settings in the previous sections, while the fixed parameters are reported in \ref{app:constants}. In order to perform a comparison purely on the Boltzmann solver, we read in the \class\ recombination files for each model and set the same initial conditions\footnote{Note that \texttt{initial\_conditions=class} in \PyCosmo\ corresponds to adiabatic initial conditions in \class.}.  All the computations are carried out on a single core on a full node of the ETH Zurich Euler cluster\footnote{Euler is a HPC Cluster of ETH Zurich, a description of the hardware in an Euler VI node can be found at \hyperlink{https://scicomp.ethz.ch/wiki/Euler}{https://scicomp.ethz.ch/wiki/Euler} (two 64-core AMD EPYC 7742 processors).}, by disabling parallel execution in \class\ and not enabling parallel computation for the \PyCosmo\ power spectrum. We run on a full node on the cluster, instead of on a laptop, in order to only run the Boltzmann solver and not get impacted by other processes being executed by the operating system. In the case of \PyCosmo, we only report the time necessary for the power spectrum computation, which does not include the compilation time for the first time the \CCpp code is generated for each model. Most models also require a second compilation (recompilation) that applies permutations to the existing equations in order to enable the use of specialized solvers instead of the standard solver as explained in Section \ref{sec:codegen}. The runtime we report is the best time obtained in three executions.

We begin by comparing the runtime between the computation of the matter power spectrum in \PyCosmo, with and without turning on the RSA, and \class, both for the \textit{speed} and \textit{precision} settings (defined in Section \ref{sec:precision}). In Table~\ref{tab:rsa_table}, we display the time necessary to compute the power spectrum with \PyCosmo\ and \class\ for a 100 log-spaced $k$ values between $k_{min}$ and $k_{max}$. We fix $k_{min}$ = $10^{-4}\ \mathrm{Mpc^{-1}}$, and set $k_{max}$ first to $1\ \mathrm{Mpc^{-1}}$ and then to $10\ \mathrm{Mpc^{-1}}$, where the effects of the Radiation Streaming Approximation are most evident.
\begin{table}
    \centering
    \begin{tabular}{|c|c|c|c|c||c|c|c|}
    \hline
    \textbf{Model} & $k_{max}$  &\multicolumn{3}{|c||}{\textit{Speed} settings, time [s] }&\multicolumn{3}{|c|}{\textit{Precision} settings, time [s] }\\
    \hline
     & & \texttt{PyCosmo} &\texttt{PyCosmo RSA} & \class\ & \texttt{PyCosmo} &  \texttt{PyCosmo RSA} & \class\ \\
    \hline
    $\Lambda$CDM & 1 $\mathrm{Mpc^{-1}}$& 1.26 & 0.23 & 0.42 & 3.80 & 1.05& 2.02\\
    \hline
    $\Lambda$CDM & 10 $\mathrm{Mpc^{-1}}$& 8.80 & 0.44& 0.80& 20.5& 2.20& 5.28\\
    \hline
    $w$CDM & 1 $\mathrm{Mpc^{-1}}$& 1.32 & 0.65& 1.29&3.84 &1.55 &2.86 \\
    \hline
    $w$CDM & 10 $\mathrm{Mpc^{-1}}$&9.08 &0.82 &4.91 &20.93 &2.72 & 10.18\\
    \hline
     degenerate $M_\nu$ & 1 $\mathrm{Mpc^{-1}}$& 54.54&29.04 &10.19 &237.26 &154.93 &105.87\\
    \hline
    degenerate $M_\nu$ & 10 $\mathrm{Mpc^{-1}}$&357.24 &98.52 &13.78 &1337.32 &471.22 &417.95\\
    \hline  
    \end{tabular}
    \caption{Best execution time from three executions on a full Euler VI node.  }
    \label{tab:rsa_table}
\end{table}

 We observe that, while \PyCosmo\ achieves a slower runtime than \class\ before introducing the RSA, the RSA reverts the situation for all the models, except for massive neutrinos. This happens despite the presence of further approximations in the \class\ implementation. Previous versions of \PyCosmo\ achieved a comparable execution time with \class\ without physical approximations \citep{pycosmo}. This was due to the reduction of the dynamic range of the time step based on a consistency relation of the Einstein equations. The adaptive control of the time step was highly optimized for \LCDM and proved difficult to extend to more general models. The new approach has also the advantage of reducing the number of permutations necessary to create the optimized code.
 
 In the case of massive neutrinos, the size of the system determines a considerable decline in the performance of both codes. \PyCosmo\ is significantly slower than \class\ when using the \textit{speed} settings, mainly due to the presence of the fluid approximation for non cold dark matter in \class. For the $precision$ settings, the time needed for the computation is comparable for \class\ and \PyCosmo.  The ODE system used by \class\ is larger in this case, since the \texttt{tol\_ncdm} parameters are set to $10^{-10}$ in \texttt{pk\_ref.pre}, and the three neutrinos are treated independently despite having the same mass. We do not decrease \texttt{mnu\_relerr}, equivalent to \texttt{tol\_ncdm}, in \PyCosmo\ since we do not deem it necessary to achieve the desired precision. The error caused by sampling less $q$ values is always subdominant compared to the hierarchy truncation, due to the small contribution of massive neutrinos to the overall matter density. 
  
 We compare the numerical results of \PyCosmo\ and \class\ in Table~\ref{tab:compare_class_prec}. We report the maximum relative difference between the matter power spectra in \texttt{PyCosmo} and \class\ in five $k$ ranges
  for the different models and precision settings. The RSA is not relevant here because it leads to a $10^{-5}$ relative difference, when compared to the full computation with the same \texttt{l\_max}, as shown for $\Lambda$CDM in the previous section.

\begin{table}[t]
\centering
\begin{tabular}{|c|c|c|c|c|c|c|}
    \hline
    \multicolumn{1}{|c|}{\texttt{Model}}& \multicolumn{1}{c|}{precision}& \multicolumn{5}{c|}{$k$ range [$\mathrm{Mpc^{-1}}$]} \\
    \hline
     & & $10^{-4} - 10^{-3}$ & $10^{-3} - 10^{-2}$ &  $10^{-2} - 10^{-1}$ & $10^{-1} - 1$ & $1 - 10$ \\
    \hline
    \hline
    \multirow{2}{*}{\textbf{$\Lambda$CDM}} & \textit{speed} &$0.0006$ & $0.0007$ & $0.0007$ & $0.0031$ & $0.0077$\\
    \cline{2-7}
     & \textit{precision} & $0.0001$ & $0.0003$ & $0.0005$ & $0.0003$ & $0.0008$\\
    \hline
    \hline
    \multirow{2}{*}{\textbf{$w$CDM}} & \textit{speed} & $0.0006$ & $0.0007$ & $0.0008$  & $0.0031$ & $0.0078$\\
    \cline{2-7}
     & \textit{precision}& $0.0002$ & $0.0003$ & $0.0005$ & $0.0003$ & $0.0008$\\
    \hline
    \hline
    \multirow{2}{*}{\textbf{degenerate $M_{\nu}$}} & \textit{speed}& $0.0004$& $0.0010$&$0.0010$ &$0.0046$ & $0.0146$\\
    \cline{2-7}
     & \textit{precision}&$0.0001$ &$0.0003$ & $0.0005$ &$0.0002$ & $0.0001$\\
     \hline
\end{tabular}
\caption{Maximum relative differences in $k$ ranges between \PyCosmo\ and \class\ for given models and precision settings.}
\label{tab:compare_class_prec}
\end{table}

We start by noting that the size of the relative differences is comparable in all models when using the same $k$ ranges, with the exception of the model with massive neutrinos for $k>0.1\ \mathrm{Mpc^{-1}}$. We also see that the difference in precision between \textit{speed} and \textit{precision} settings is dominant on scales $k>0.1\ \mathrm{Mpc^{-1}}$, where the truncation effects have the largest impact. The \textit{precision} settings lead to a relative difference to \class\ that is better than $0.1\%$ on all scales $k < 10\ \mathrm{Mpc^{-1}}$, while the \textit{speed} settings lead to a relative difference of order $0.5-1\%$ beyond $k=0.1\ \mathrm{Mpc^{-1}}$. This is acceptable, especially since \texttt{cl\_permille.pre} comes with no guarantees for scales $k>1\ h\mathrm{Mpc^{-1}}$.

\section{Conclusion}
\label{sec:conclusion}
In this paper, we demonstrated how the \PyCosmo\ Boltzmann solver can be easily modified to include extensions of the $\Lambda$CDM cosmological model and approximation schemes, by taking advantage of the \sympy symbolic implementation of equations. The symbolic expressions are translated into optimized \CCpp code by the \sympytoc package presented in \cite{uwe}. In this way, \PyCosmo\ combines the speed of \CCpp with the user-friendliness of symbolic \texttt{Python}.

We first presented two cosmological model extensions: dark energy with a constant equation of state, which is a minimal modification of $\Lambda$CDM, and massive neutrinos, which enlarge considerably the ODE system and comprise numerical integrations, constituting a more complex extension. The inclusion of these models makes \PyCosmo\ more widely applicable for constraining cosmology. 
We also implemented an approximation scheme, the radiation streaming approximation. In order to trigger the approximation, \sympytoc includes a functionality to switch between two different ODE systems when a condition is verified.  The radiation streaming approximation makes the solution of the ODE system considerably faster, up to an 80\% speed-up, since it suppresses oscillations of the radiation fields for large $k$ values. The errors introduced by the approximation are largely sub-dominant compared to the artificial power reflection induced by the hierarchy truncation. For convenience, we presented a conversion table between common conventions for cosmological perturbations (\ref{app:notation}) and a clarification of the computation of the total matter power spectrum with $A_s$ normalization (\ref{app:pk}).

We compared the numerical results obtained by computing the total matter power spectrum with \class\ and found an agreement better than 0.1\% with high precision settings. With more relaxed precision settings, we found an agreement of 0.5\% for scales $k<1\mathrm{Mpc^{-1}}$ for all models. The \PyCosmo\ Boltzmann solver achieves precision and speed that is comparable to \class, while not relying on physical approximations (such as tight coupling and ultra relativistic fluid approximation) other than the radiation streaming approximation introduced in this work. In the future, we plan to include more beyond $\Lambda$CDM models in \PyCosmo. Possible extensions include  time-varying dark energy, early dark energy \citep{Poulin:2018cxd,Hill:2020osr}, curvature \citep{Pitrou:2020lhu}, dark matter models \citep{Hui:2021tkt}, such as axions and fuzzy dark matter, and extensions of the neutrino sector. We believe that our symbolic implementation will be applicable for most model extensions, with some refactoring needed when a model introduces an ODE system already at background level (for example in the case of scalar field models).

\section{Acknowledgements}
This work was supported in part by grant 200021\_192243 from the Swiss National Science Foundation. The authors thank Thomas Tram for answering questions about the \texttt{CLASS} implementation. We thank Joel Mayor for his comments that led to some corrections and clean up of the code, and Silvan Fischbacher for helpful discussions. The authors also thank Joel Akeret, Adam Amara, Lukas Gamper, Jörg Herbel, Tomasz Kacprzak and Andrina Nicola for discussions and contributions to earlier versions of \PyCosmo. In this work, we rely on the \texttt{Python} packages \texttt{numpy}
\citep{numpy}, \texttt{scipy} \citep{scipy}, \texttt{matplotlib} \citep{matplotlib} and \texttt{SymPy} \citep{sympy}.

\clearpage

\begin{appendices}
\section{Notation table}
\label{app:notation}
\begin{table}[h!]
{\setlength{\tabcolsep}{8pt}
\resizebox{0.99\textwidth}{!}{
\begin{tabular}{ccc}
Ma-Bertschinger & \PyCosmo & Meaning \\
\hline
$a$ & $a$ & scale factor\\
$k$ & $k$ & wavenumber of Fourier mode\\
$P_i$ & $P_i$ & conjugate momentum to $x_i$\\
$p_i$ & $p_i$ & proper momentum \\
$q_i$ & $q_i$ & $a p_i$\\
$\epsilon_i$ & $\epsilon_i$ & $(q_i^2 + am^2)^{1/2}$\\
$\delta_{\gamma} $& $4 \Theta_0$ & photons overdensity \\
$\theta_{\gamma}$& $3 k \Theta_1$ & photons velocity divergence\\
$\sigma_{\gamma}$ &$2 \Theta_2$ & photons shear stress\\
$F_{\gamma l}$ & $4\Theta_l$ & $l$th Legendre component of photons perturbations\\
$\delta_{ur}$ & $4\mathcal{N}_0$ & massless neutrinos overdensity \\
$\theta_{ur}$ & $3 k \mathcal{N}_1$ & divergence of massless neutrinos velocity\\
$\sigma_{ur}$ & $2 \mathcal{N}_2$ & shear stress of massless neutrino fluid\\
$\psi, \phi$ & $\Psi, -\Phi$ & Newtonian gravitational potentials\\
$\tau$&$\eta$ &conformal time\\
$\tau_c$=1/$\kappa'$ & $-\dot{\tau}$ & Thomson scattering rate \\
$\theta_c$ & $k u = ik v$ & velocity divergence of dark matter\\
$\theta_b$ & $k u_b = ik v_b$ & velocity divergence of baryons\\
$\Psi_1,\Psi_2, \Psi_l$ & $\mathcal{M}_1, \mathcal{M}_2, \mathcal{M}_l$ & Legendre components of massive neutrinos perturbations\\
$\delta_{h}$ & $\delta_{\nu,m}$ & massive neutrinos overdensity \\
$\theta_h $ & $k u_{\nu,m}$ & massive neutrinos velocity divergence\\
$\sigma_h$ & $\sigma_{\nu,m}$ & massive neutrinos shear stress\\
\end{tabular}}}
\vspace{10pt}
\caption{Relations between the notation in Ma-Bertschinger and \PyCosmo. Ma-Bertschinger notation is used in both the Boltzmann solvers \texttt{COSMICS} and \class. Some exceptions include $\theta_{ur}$ that is denoted $\theta_r$ in \texttt{COSMICS} and all the massive neutrino quantities that are denoted $\delta_{ncdm},\ \theta_{ncdm}$ and $\sigma_{ncdm}$ in \class\ (\texttt{ncdm} = non cold dark matter) and $\delta_n,\ \theta_n$ and $\sigma_n$ in \texttt{COSMICS}.}
\end{table}
\newpage

\section{Linear perturbations in $\ln(a)$}
\label{app:eqns}
In \texttt{PyCosmo} we use $\ln(a)$ (from here on simply $lna$) as the independent variable of the Einstein-Boltzmann ODE system. The relation between $\ln(a)$ and $\eta$ is simply: $\frac{d\ln(a)}{d\eta}= aH$. We report the Einstein-Boltzmann equations in $\Lambda$CDM, as used in the code, in the following: 
\begin{itemize}
\item Einstein equations:
\begin{align*}
\Psi &= -\Phi - 12 \left(\frac{H_0}{k a}\right)^2 \left(\Omega_{\gamma} \Theta_2 + \Omega_{\nu} \mathcal{N}_2\right)\\
\Pi &= \Theta_2 + \Theta_{P,0} + \Theta_{P,2}\\
\frac{d\Phi}{dlna} &= \Psi - \left(\frac{k}{aH}\right)^2 \frac{\Phi}{3} + \frac{1}{2}\left(\frac{H_0}{H}\right)^2 \left(\left(\Omega_{dm} \delta + \Omega_b \delta_b\right) a^{-3} + 4 a^{-4} \left( \Omega_{\gamma} \Theta_0 + \Omega_{\nu} \mathcal{N}_0 \right) \right)
\end{align*}
\item Dark matter:
\begin{align*}
\frac{d\delta}{dlna} &= -\frac{k}{aH} u - 3 \frac{d\Phi}{dlna}\\
\frac{du}{dlna} &= -u + \frac{k}{aH}\Psi
\end{align*}
\item Baryonic matter:
\begin{align*}
\frac{d \delta_b}{dlna} &= -\frac{k}{aH} u_b - 3 \frac{d\Phi}{dlna}\\
\frac{du_b}{dlna} &= -u_b + \frac{k}{aH} \Psi + \frac{\dot{\tau}}{R a H} \left( u_b - 3 \Theta_1 \right) + \frac{k}{aH} {c_s}^2 \delta_b 
\end{align*}
\item Photons temperature:
\begin{align*}
\frac{d\Theta_0}{dlna} &= -\frac{k}{aH} \Theta_1 -\frac{d\Phi}{dlna} \\
\frac{d\Theta_1}{dlna} &= \frac{k}{3aH} \left(\Theta_0 - 2 \Theta_2 + \Psi \right) + \frac{\dot{\tau}}{aH}\left(\Theta_1 -\frac{u_b}{3}\right)\\
\frac{d\Theta_2}{dlna} &= \frac{k}{5aH} \left( 2 \Theta_1 - 3 \Theta_3 \right) + \frac{\dot{\tau}}{aH} \left(\Theta_{P,0} - \frac{\Pi}{2} \right) 
\end{align*}
For $l > 2$
$$\frac{d\Theta_l}{dlna} = \frac{k}{aH(2l + 1)} \left(l \Theta_{l-1}
- (l+1) \Theta_{l+1}\right) + \frac{\dot{\tau}}{aH}\Theta_l $$
\item Photons polarization:
\begin{align*}
\frac{d\Theta_{P,0}}{dlna} &= -\frac{k}{aH}\Theta_{P,1} + \frac{\dot{\tau}}{aH}\left(\Theta_{P,0} - \frac{\Pi}{2}\right)\\
\frac{d\Theta_{P,1}}{dlna} &= \frac{k}{3aH} \left( \Theta_{P,0} - 2\Theta_{P,2}\right) + \frac{\dot{\tau}}{aH}\Theta_{P,1}\\
\frac{d\Theta_{P,2}}{dlna} &= \frac{k}{5aH} \left( 2\Theta_{P,1} - 3\Theta_{P,3}\right) + \frac{\dot{\tau}}{aH}\left(\Theta_{P,2} - \frac{\Pi}{10}\right)
\end{align*}
For $l > 2$
$$\frac{d\Theta_{P,l}}{dlna} = \frac{k}{aH(2l + 1)} \left( l\Theta_{P,l-1} - (l+1)\Theta_{P,l+1}\right) + \frac{\dot{\tau}}{aH} \Theta_{P,l}$$
\item Massless neutrinos:
\begin{align*}
\frac{d\mathcal{N}_0}{dlna} &= -\frac{k}{aH} \mathcal{N}_1 - \frac{d\Phi}{dlna}\\
\frac{d\mathcal{N}_1}{dlna} &= \frac{k}{3aH} \left(\mathcal{N}_0-2 \mathcal{N}_2 + \Psi\right)
\end{align*}
For $l \geq 2$
$$\frac{d\mathcal{N}_l}{dlna} = \frac{k}{(2l + 1) aH} \left( l \mathcal{N}_{l-1} - (l+1) \mathcal{N}_{l+1}\right)$$
The hierarchy truncations for relativistic species read:
\begin{align*}
\frac{d\Theta_{l_{max}}}{dlna} &= \frac{1}{aH}\left(k \Theta_{l_{max}-1} - \left(\frac{(l_{max}+1)}{\eta}- \dot{\tau}\right) \Theta_{l_{max}}\right) \\
\frac{d\Theta_{P,l_{max}}}{dlna} &= \frac{1}{aH}\left( k \Theta_{P,l_{max}} - \left(\frac{(l_{max}+1)}{\eta} - \dot{\tau}\right) \Theta_{P,l_{max}}\right) \\
\frac{d\mathcal{N}_{l_{max}}}{dlna} &= \frac{1}{aH} \left( k \mathcal{N}_{l_{max}-1} - \frac{l_{max}+1}{\eta} \mathcal{N}_{l_{max}} \right) 
\end{align*}
\end{itemize}
The equations for the extended models are easily obtained with the same change of variables from the equations in $\eta$, reported in the corresponding sections of the paper.
\section{Adiabatic initial conditions}
\label{app:ini}
The default setting of \texttt{PyCosmo} is to use the adiabatic initial conditions from \texttt{CLASS} \citep{Blas_2011,Cyr_Racine_2011, Bucher_2000}. We present the initial conditions for $\Lambda$CDM, wCDM and massive neutrinos in the following.
We start by introducing auxiliary notation which is useful to define the initial conditions:

\begin{align*}
\eta_0 &= \eta(a_0)\hspace{3.3 cm}
 \frac{\dot{a}}{a} &= aH(a_0) \hspace{2cm}
\omega &= \frac{\Omega_m H_0}{\sqrt{\Omega_{r,tot}}}\hspace{2cm} \frac{\rho_m}{\rho_r}&=\frac{\Omega_m a_0}{\Omega_r}\\
F_{\nu} &= \frac{\Omega_{\nu,tot}}{\Omega_{r,tot}} \hspace{2.8 cm}
F_{cdm} &= \frac{\Omega_{cdm}}{\Omega_m}\hspace{2 cm}
F_b &= 1-F_{cdm}\hspace{2 cm}
F_g &= 1-F_{\nu}\\
\end{align*}

where $\Omega_{\nu,tot} =  \frac{7}{8}N_\mathrm{eff}\left(\frac{4}{11}\right)^{\frac{4}{3}}\Omega_\gamma$ which includes massless and massive neutrinos (assumed to be relativistic at early times), $\Omega_{r,tot}=\Omega_\gamma + \Omega_{\nu,tot}$  and $\eta_0$ is defined as the minimum between $0.001/k$ and $0.1h$ with $k$ expressed in $h$Mpc$^{-1}$. $a_0$ is computed from $\eta_0$ assuming radiation domination as $a_0 = \eta_0 \sqrt{\Omega_{r,tot}}/ H_0$. We use these conservative definitions of initial times to avoid using an iterative shooting algorithm. 
The initial perturbations are computed in synchronous gauge as
\begin{center}
    \begin{align*}
    \delta_g &= - \frac{(k\eta_0)^2(1-\frac{\omega\eta_0}{5})}{3}\\
    \theta_g &= -\frac{k(k\eta_0)^3}{36} \left( 1-\frac{3(1+5F_b-F_{\nu})}{20(1-F_{\nu})} \omega \eta_0 \right)\\
    \delta_b &= \frac{3\delta_g}{4} = \delta_c\\
    \theta_b &= \theta_g\\
    \delta_{ur} &= \delta_g \\
    \theta_{ur} &= - \frac{k(k\eta_0)^3}{36(4F_{\nu}+15)} \left(4 F_{\nu}+23- \frac{3(8F_{\nu}^2+50 F_{\nu}+275)}{20(2 F_{\nu} + 15) } \eta_0 \omega\right)\\
    \sigma_{ur} &= \frac{2(k\eta_0)^2}{45+12F_{\nu}} \left(1+\frac{(4F_{\nu}-5)}{4(2F_{\nu}+15)} \eta_0 \omega\right)\\
    l3_{ur} &= \frac{2(k\eta_0)^3}{7(12F_{\nu}+45)}\\
    \eta_{sync} &= 1-\frac{(k\eta_0)^2}{12(15+4F_{\nu})} (5+4F_{\nu} - \frac{(16F_{\nu}^2 + 280 F_{\nu} + 325)}{10(2F_{\nu}+15)}\eta_0 \omega).
\end{align*}
\end{center}

In the wCDM case we add the generalized initial adiabatic conditions from \cite{Ballesteros:2010ks} for the dark energy fields:
\begin{align*}
    \delta_\mathrm{de}&= -\frac{1}{4}(1+w_\mathrm{de})\frac{4-3c_\mathrm{s,de}^2}{4-6w_\mathrm{de}+3c_\mathrm{s,de}^2}(k\eta_0)^2\\
    \theta_\mathrm{de}&= -\frac{kc_\mathrm{s,de}^2}{4(4-6w_\mathrm{de}+3c_\mathrm{s,de}^2)}(k\eta_0)^3.
    \end{align*}

We then need to introduce the transformation from synchronous to conformal Newtonian gauge that uses the following quantities
\begin{align*}
\delta_{tot} &= \frac{F_g\delta_g+F_{\nu}\delta_{ur}+\frac{\rho_m}{\rho_r}(F_b\delta_b+F_{cdm}\delta_c)}{1+\frac{\rho_m}{\rho_r}}\\
v_{tot} &= \frac{\frac{4}{3}(F_g\theta_g+F_{\nu}\theta_{ur}) + \frac{\rho_m}{\rho_r}F_b\theta_b}{(
1+\frac{\rho_m}{\rho_r})}\\
\alpha &= \frac{\eta_{sync} + \frac{3}{2}	\left(\frac{\dot{a}}{a}\right)^2\frac{1}{k^2}(\delta_{tot} + 3\frac{\dot{a}}{a}\frac{v_{tot}}{k^2})}{\frac{\dot{a}}{a}}
\end{align*}
 
and reads
\begin{align*}
-\Phi &= \eta_{sync}-\frac{\dot{a}}{a}	\alpha \\
\delta_c &= \delta_c(sync) - 3\frac{\dot{a}}{a}\alpha &
\theta_c &= k^2\alpha \\
\delta_b &= \delta_b(sync)-3\frac{\dot{a}}{a}\alpha &
\theta_b &= \theta_b(sync)+k^2 \alpha \\
\delta_g &= \delta_g(sync) - 4\frac{\dot{a}}{a}\alpha &
\theta_g &= \theta_g(sync)+k^2\alpha \\
\delta_{ur} &= \delta_{ur}(sync)-4\frac{\dot{a}}{a} &
\theta_{ur} &= \theta_{ur}(sync) + k^2\alpha  \\
\delta_{\mathrm{de}} &= \delta_{\mathrm{de}}(sync) -3(1+w_{\mathrm{de}}) \frac{\dot{a}}{a} \alpha &
\theta_{\mathrm{de}} &= \theta_{\mathrm{de}}(sync) + k^2\alpha.
\end{align*}
The fields where we did not specify a conversion are gauge invariant, including the massive neutrinos' perturbations of the distribution function. The initial conditions imposed to the multipoles of $\mathcal{M}(\mu)$ are

\begin{align*}
\mathcal{M}_0 &= -\frac{1}{4}\delta_{\nu} \frac{dlnf_0}{d\ln{q}} = \frac{q}{k_b T_0}\frac{\mathcal{N}_0}{e^{-q/k_b T_0}+1}\\ 
\mathcal{M}_1 &= -\frac{\epsilon}{3qk}\theta_{\nu} \frac{dlnf_0}{d\ln{q}} = \frac{\epsilon}{k_b T_0} \frac{\mathcal{N}_1}{e^{-q/k_b T_0}+1}\\
\mathcal{M}_2 &= -\frac{1}{2}\sigma_{\nu}\frac{dlnf_0}{d\ln{q}} = \frac{q}{k_b T_0}\frac{\mathcal{N}_2}{e^{-q/k_bT_0}+1}\\
\mathcal{M}_3 &= -\frac{1}{4}l3_{ur} \frac{dlnf_0}{d\ln{q}} =  \frac{q}{k_b T_0}\frac{\mathcal{N}_3}{e^{-q/k_bT_0}+1}
\end{align*}

and are set after transforming the other fields to the conformal Newtonian gauge. We can then relate the fields from \texttt{CLASS} to those of \texttt{PyCosmo} with the following conversions (already introduced in Appendix~\ref{app:notation}):

\begin{align*}
u &= \theta_c/k &
u_b &= \theta_b/k &
u_\mathrm{de} &= \theta_\mathrm{de}/k &
\Theta_0 &= \delta_g / 4 &
\mathcal{N}_0 &= \delta_{ur}/4 \\
\Theta_1 &= \theta_g / 3k &
\mathcal{N}_1 &= \theta_{ur}/3k &
\mathcal{N}_2 &= \sigma_{ur}/2 &
\mathcal{N}_3 &= l3_{ur}/4.
\end{align*}
\section{Fixed cosmological parameters}
\label{app:constants}
We report all the parameters of the configuration file of \texttt{PyCosmo} that are kept fixed throughout the paper:
\begin{multicols}{2}

\begin{verbatim}
[cosmology]
h = 0.7
omega_b = 0.06
flat_universe = True
Tcmb = 2.725
Yp = 0.24
wa = 0.0
cs_de2 = 1.0
T_mnu = 0.71611

[recombination]
recomb = `class'

[linear_perturbations]
pk_type = `boltz'
pk_norm_type = `A_s'
pk_norm = 2.1e-9
k_pivot = 0.05

[internal:boltzmann_solver]
initial_conditions = `class'
dt_0 = 1.5e-2
sec_factor = 10.0
boltzmann_max_bdf_order = 5
boltzmann_max_iter = 10000000
fast_solver = True

[internal:physical_constants]
kb = 8.617342790900664e-05
evc2 = 1.7826617580683397e-36
G = 6.67428e-11
hbar = 6.582118991312934e-16
mpc = 3.085677581282e22
mp = 938.272013425824
msun = 1.98855e30
sigmat = 6.6524616e-29
\end{verbatim}
\end{multicols}
Note that the physical constants are not set to the default values in \PyCosmo, but to default values from \class\ (found in the header files \texttt{thermodynamics.h} and \texttt{background.h}). 

The same parameters are passed to \class\ and are also fixed throughout this work:
\begin{multicols}{2}
\begin{verbatim}
output = `mPk'
T_cmb = 2.725
Omega_b = 0.06
h = 0.7
T_ncdm =  0.71611, 0.71611, 0.71611
ksi_ncdm = 0, 0, 0
Omega_fld = 0
wa_fld = 0
cs2_fld = 1
reio_parametrization = `reio_none'
YHe = 0.24
gauge = `newtonian'
A_s = 2.1e-09
n_s = 1
alpha_s = 0
k_pivot = 0.05
\end{verbatim}
\end{multicols}
Note that some parameters (for instance $T_{\nu,m}$ or $c^2_{s,de}$) are specified only when necessary.

\section{Power spectrum computation}
\label{app:pk}
We compute the gauge invariant real space matter power spectrum, using a fully general relativistic treatment \citep{Yoo_2009, Yoo_2010, Bonvin_2011, Challinor_2011}, accounting for real space matter fluctuations and volume distortions, similarly to  \class\ \citep{Dio_2013}. Note that general relativistic corrections are not included in other observables in \PyCosmo, but are left as future development. Prior versions of \PyCosmo\ separated transfer function and growth factor and used the Poisson equation to relate the Newtonian gauge matter density perturbation to the Newtonian gravitational potential. This is still the case when setting \texttt{pk\_norm\_type} to \texttt{deltah} and using the power spectrum fitting functions.

We define $\Omega_{m,tot} = \Omega_{dm} + \Omega_b + \Omega_{\nu,m}$ as the total matter energy density, including massive neutrinos and $P_{m,tot}=  P_{\nu,m}$, since the massive neutrino component is the only matter ingredient that has a non-zero pressure term. Then the gauge invariant matter density reads \begin{equation}
\begin{split}
   \delta_{m,tot} = \frac{\delta\rho_{m,tot}}{\bar{\rho}_{m,tot}} + 3 \frac{aH}{k^2}\theta_{m,tot} =\ & \frac{\Omega_{dm}\delta + \Omega_b \delta_b + \Omega_{\nu,m}\delta_{\nu,m}}{\Omega_{m,tot}} \\
   & + 3 \frac{aH}{k} \frac{\Omega_{dm}u  + \Omega_b u_b + (\Omega_{\nu,m} + P_{\nu,m})u_{\nu,m}}{\Omega_{m,tot}+P_{m,tot}}, 
\end{split}
\end{equation} 
where we omitted the $a$ dependencies for brevity.

The power spectrum is defined in terms of the primordial power spectrum of gauge invariant curvature perturbations, $P_\mathcal{R}(k) = \mathcal{A}_s \left(\frac{k}{k_p}\right)^{n_s-1}$ with $n_s$ the tilt of the primordial power spectrum and $k_p$ the pivot scale with corresponding $\mathcal{A}_s$ amplitude , as
\begin{equation}
    P(k,a) = 2\pi^2 \mathcal{A}_s \frac{k^{n_s - 4}}{k_p^{n_s-1}} \delta_{m,tot}(k,a)^2,
\end{equation}
valid in the case of adiabatic initial conditions for which the initial curvature $\mathcal{R}$ is normalized to 1. These are the only initial conditions currently implemented in \PyCosmo. \PyCosmo\ also allows to output $P_{cb}(k,a)$, the power spectrum of dark and baryonic matter, where massive neutrinos are excluded.
\end{appendices}

\bibliography{pycosmo}

\begin{thebibliography}{50}
\expandafter\ifx\csname natexlab\endcsname\relax\def\natexlab#1{#1}\fi
\providecommand{\url}[1]{\texttt{#1}}
\providecommand{\href}[2]{#2}
\providecommand{\path}[1]{#1}
\providecommand{\DOIprefix}{doi:}
\providecommand{\ArXivprefix}{arXiv:}
\providecommand{\URLprefix}{URL: }
\providecommand{\Pubmedprefix}{pmid:}
\providecommand{\doi}[1]{\href{http://dx.doi.org/#1}{\path{#1}}}
\providecommand{\Pubmed}[1]{\href{pmid:#1}{\path{#1}}}
\providecommand{\bibinfo}[2]{#2}
\ifx\xfnm\relax \def\xfnm[#1]{\unskip,\space#1}\fi
\bibitem[{Abbott et~al.(2021)}]{Abbott:2021bzy}
\bibinfo{author}{Abbott, T. M.~C.} et~al. (\bibinfo{collaboration}{DES})
  (\bibinfo{year}{2021}).
\newblock \bibinfo{title}{{Dark Energy Survey Year 3 Results: Cosmological
  Constraints from Galaxy Clustering and Weak Lensing}}, .
\newblock \href{http://arxiv.org/abs/2105.13549}{\tt arXiv:2105.13549}.
\bibitem[{Aghanim et~al.(2020)}]{Aghanim:2018eyx}
\bibinfo{author}{Aghanim, N.} et~al. (\bibinfo{collaboration}{Planck})
  (\bibinfo{year}{2020}).
\newblock \bibinfo{title}{{Planck 2018 results. VI. Cosmological parameters}}.
\newblock {\it \bibinfo{journal}{Astron. Astrophys.}\/},  {\it
  \bibinfo{volume}{641}\/}, \bibinfo{pages}{A6}.
  \DOIprefix\doi{10.1051/0004-6361/201833910}.
  \href{http://arxiv.org/abs/1807.06209}{\tt arXiv:1807.06209}.
\bibitem[{Ahmad et~al.(2002)}]{PhysRevLett.89.011301}
\bibinfo{author}{Ahmad, Q.~R.} et~al. (\bibinfo{collaboration}{SNO
  Collaboration}) (\bibinfo{year}{2002}).
\newblock \bibinfo{title}{Direct evidence for neutrino flavor transformation
  from neutral-current interactions in the sudbury neutrino observatory}.
\newblock {\it \bibinfo{journal}{Phys. Rev. Lett.}\/},  {\it
  \bibinfo{volume}{89}\/}, \bibinfo{pages}{011301}. \URLprefix
  \url{https://link.aps.org/doi/10.1103/PhysRevLett.89.011301}.
  \DOIprefix\doi{10.1103/PhysRevLett.89.011301}.
\bibitem[{Ballesteros \& Lesgourgues(2010)}]{Ballesteros:2010ks}
\bibinfo{author}{Ballesteros, G.}, \& \bibinfo{author}{Lesgourgues, J.}
  (\bibinfo{year}{2010}).
\newblock \bibinfo{title}{{Dark energy with non-adiabatic sound speed: initial
  conditions and detectability}}.
\newblock {\it \bibinfo{journal}{JCAP}\/},  {\it \bibinfo{volume}{10}\/},
  \bibinfo{pages}{014}. \DOIprefix\doi{10.1088/1475-7516/2010/10/014}.
  \href{http://arxiv.org/abs/1004.5509}{\tt arXiv:1004.5509}.
\bibitem[{{Bertschinger}(1995)}]{Bertschinger_1995}
\bibinfo{author}{{Bertschinger}, E.} (\bibinfo{year}{1995}).
\newblock \bibinfo{title}{{COSMICS: Cosmological Initial Conditions and
  Microwave Anisotropy Codes}}.
\newblock {\it \bibinfo{journal}{arXiv e-prints}\/},  (pp.
  \bibinfo{pages}{astro--ph/9506070}).
  \href{http://arxiv.org/abs/astro-ph/9506070}{\tt arXiv:astro-ph/9506070}.
\bibitem[{Blas et~al.(2011)Blas, Lesgourgues \& Tram}]{Blas_2011}
\bibinfo{author}{Blas, D.}, \bibinfo{author}{Lesgourgues, J.}, \&
  \bibinfo{author}{Tram, T.} (\bibinfo{year}{2011}).
\newblock \bibinfo{title}{The cosmic linear anisotropy solving system (class).
  part ii: Approximation schemes}.
\newblock {\it \bibinfo{journal}{Journal of Cosmology and Astroparticle
  Physics}\/},  {\it \bibinfo{volume}{2011}\/}, \bibinfo{pages}{034–034}.
  \URLprefix \url{http://dx.doi.org/10.1088/1475-7516/2011/07/034}.
  \DOIprefix\doi{10.1088/1475-7516/2011/07/034}.
\bibitem[{Bonvin \& Durrer(2011)}]{Bonvin_2011}
\bibinfo{author}{Bonvin, C.}, \& \bibinfo{author}{Durrer, R.}
  (\bibinfo{year}{2011}).
\newblock \bibinfo{title}{What galaxy surveys really measure}.
\newblock {\it \bibinfo{journal}{Physical Review D}\/},  {\it
  \bibinfo{volume}{84}\/}. \URLprefix
  \url{http://dx.doi.org/10.1103/PhysRevD.84.063505}.
  \DOIprefix\doi{10.1103/physrevd.84.063505}.
\bibitem[{Bucher et~al.(2000)Bucher, Moodley \& Turok}]{Bucher_2000}
\bibinfo{author}{Bucher, M.}, \bibinfo{author}{Moodley, K.}, \&
  \bibinfo{author}{Turok, N.} (\bibinfo{year}{2000}).
\newblock \bibinfo{title}{General primordial cosmic perturbation}.
\newblock {\it \bibinfo{journal}{Physical Review D}\/},  {\it
  \bibinfo{volume}{62}\/}. \URLprefix
  \url{https://doi.org/10.1103%2Fphysrevd.62.083508}.
  \DOIprefix\doi{10.1103/physrevd.62.083508}.
\bibitem[{Challinor \& Lewis(2011)}]{Challinor_2011}
\bibinfo{author}{Challinor, A.}, \& \bibinfo{author}{Lewis, A.}
  (\bibinfo{year}{2011}).
\newblock \bibinfo{title}{Linear power spectrum of observed source number
  counts}.
\newblock {\it \bibinfo{journal}{Physical Review D}\/},  {\it
  \bibinfo{volume}{84}\/}. \URLprefix
  \url{http://dx.doi.org/10.1103/PhysRevD.84.043516}.
  \DOIprefix\doi{10.1103/physrevd.84.043516}.
\bibitem[{Chisari et~al.(2019)}]{LSSTDarkEnergyScience:2018yem}
\bibinfo{author}{Chisari, N.~E.} et~al. (\bibinfo{collaboration}{LSST Dark
  Energy Science}) (\bibinfo{year}{2019}).
\newblock \bibinfo{title}{{Core Cosmology Library: Precision Cosmological
  Predictions for LSST}}.
\newblock {\it \bibinfo{journal}{Astrophys. J. Suppl.}\/},  {\it
  \bibinfo{volume}{242}\/}, \bibinfo{pages}{2}.
  \DOIprefix\doi{10.3847/1538-4365/ab1658}.
  \href{http://arxiv.org/abs/1812.05995}{\tt arXiv:1812.05995}.
\bibitem[{Cyr-Racine \& Sigurdson(2011)}]{Cyr_Racine_2011}
\bibinfo{author}{Cyr-Racine, F.-Y.}, \& \bibinfo{author}{Sigurdson, K.}
  (\bibinfo{year}{2011}).
\newblock \bibinfo{title}{Photons and baryons before atoms: Improving the
  tight-coupling approximation}.
\newblock {\it \bibinfo{journal}{Physical Review D}\/},  {\it
  \bibinfo{volume}{83}\/}. \URLprefix
  \url{https://doi.org/10.1103%2Fphysrevd.83.103521}.
  \DOIprefix\doi{10.1103/physrevd.83.103521}.
\bibitem[{Dio et~al.(2013)Dio, Montanari, Lesgourgues \& Durrer}]{Dio_2013}
\bibinfo{author}{Dio, E.~D.}, \bibinfo{author}{Montanari, F.},
  \bibinfo{author}{Lesgourgues, J.}, \& \bibinfo{author}{Durrer, R.}
  (\bibinfo{year}{2013}).
\newblock \bibinfo{title}{The classgal code for relativistic cosmological large
  scale structure}.
\newblock {\it \bibinfo{journal}{Journal of Cosmology and Astroparticle
  Physics}\/},  {\it \bibinfo{volume}{2013}\/}, \bibinfo{pages}{044–044}.
  \URLprefix \url{http://dx.doi.org/10.1088/1475-7516/2013/11/044}.
  \DOIprefix\doi{10.1088/1475-7516/2013/11/044}.
\bibitem[{Dodelson(2003)}]{dodelson}
\bibinfo{author}{Dodelson, S.} (\bibinfo{year}{2003}).
\newblock {\it \bibinfo{title}{Modern Cosmology}\/}.
\newblock \bibinfo{publisher}{Academic Press, Elsevier Science}.
\bibitem[{Doran(2005{\natexlab{a}})}]{Doran_2005}
\bibinfo{author}{Doran, M.} (\bibinfo{year}{2005}{\natexlab{a}}).
\newblock \bibinfo{title}{Cmbeasy: an object oriented code for the cosmic
  microwave background}.
\newblock {\it \bibinfo{journal}{Journal of Cosmology and Astroparticle
  Physics}\/},  {\it \bibinfo{volume}{2005}\/}, \bibinfo{pages}{011–011}.
  \URLprefix \url{http://dx.doi.org/10.1088/1475-7516/2005/10/011}.
  \DOIprefix\doi{10.1088/1475-7516/2005/10/011}.
\bibitem[{Doran(2005{\natexlab{b}})}]{Doran:2005ep}
\bibinfo{author}{Doran, M.} (\bibinfo{year}{2005}{\natexlab{b}}).
\newblock \bibinfo{title}{{Speeding up cosmological Boltzmann codes}}.
\newblock {\it \bibinfo{journal}{JCAP}\/},  {\it \bibinfo{volume}{06}\/},
  \bibinfo{pages}{011}. \DOIprefix\doi{10.1088/1475-7516/2005/06/011}.
  \href{http://arxiv.org/abs/astro-ph/0503277}{\tt arXiv:astro-ph/0503277}.
\bibitem[{Eguchi et~al.(2003)}]{PhysRevLett.90.021802}
\bibinfo{author}{Eguchi, K.} et~al. (\bibinfo{collaboration}{KamLAND
  Collaboration}) (\bibinfo{year}{2003}).
\newblock \bibinfo{title}{First results from kamland: Evidence for reactor
  antineutrino disappearance}.
\newblock {\it \bibinfo{journal}{Phys. Rev. Lett.}\/},  {\it
  \bibinfo{volume}{90}\/}, \bibinfo{pages}{021802}. \URLprefix
  \url{https://link.aps.org/doi/10.1103/PhysRevLett.90.021802}.
  \DOIprefix\doi{10.1103/PhysRevLett.90.021802}.
\bibitem[{Eisenstein \& Hu(1998)}]{Eisenstein_1998}
\bibinfo{author}{Eisenstein, D.~J.}, \& \bibinfo{author}{Hu, W.}
  (\bibinfo{year}{1998}).
\newblock \bibinfo{title}{Baryonic features in the matter transfer function}.
\newblock {\it \bibinfo{journal}{The Astrophysical Journal}\/},  {\it
  \bibinfo{volume}{496}\/}, \bibinfo{pages}{605–614}. \URLprefix
  \url{http://dx.doi.org/10.1086/305424}. \DOIprefix\doi{10.1086/305424}.
\bibitem[{Fukuda et~al.(1998)}]{PhysRevLett.81.1562}
\bibinfo{author}{Fukuda, Y.} et~al. (\bibinfo{collaboration}{Super-Kamiokande
  Collaboration}) (\bibinfo{year}{1998}).
\newblock \bibinfo{title}{Evidence for oscillation of atmospheric neutrinos}.
\newblock {\it \bibinfo{journal}{Phys. Rev. Lett.}\/},  {\it
  \bibinfo{volume}{81}\/}, \bibinfo{pages}{1562--1567}. \URLprefix
  \url{https://link.aps.org/doi/10.1103/PhysRevLett.81.1562}.
  \DOIprefix\doi{10.1103/PhysRevLett.81.1562}.
\bibitem[{Hill et~al.(2020)Hill, McDonough, Toomey \& Alexander}]{Hill:2020osr}
\bibinfo{author}{Hill, J.~C.}, \bibinfo{author}{McDonough, E.},
  \bibinfo{author}{Toomey, M.~W.}, \& \bibinfo{author}{Alexander, S.}
  (\bibinfo{year}{2020}).
\newblock \bibinfo{title}{{Early dark energy does not restore cosmological
  concordance}}.
\newblock {\it \bibinfo{journal}{Phys. Rev. D}\/},  {\it
  \bibinfo{volume}{102}\/}, \bibinfo{pages}{043507}.
  \DOIprefix\doi{10.1103/PhysRevD.102.043507}.
  \href{http://arxiv.org/abs/2003.07355}{\tt arXiv:2003.07355}.
\bibitem[{Hu et~al.(2014)Hu, Raveri, Frusciante \& Silvestri}]{Hu:2013twa}
\bibinfo{author}{Hu, B.}, \bibinfo{author}{Raveri, M.},
  \bibinfo{author}{Frusciante, N.}, \& \bibinfo{author}{Silvestri, A.}
  (\bibinfo{year}{2014}).
\newblock \bibinfo{title}{{Effective Field Theory of Cosmic Acceleration: an
  implementation in CAMB}}.
\newblock {\it \bibinfo{journal}{Phys. Rev. D}\/},  {\it
  \bibinfo{volume}{89}\/}, \bibinfo{pages}{103530}.
  \DOIprefix\doi{10.1103/PhysRevD.89.103530}.
  \href{http://arxiv.org/abs/1312.5742}{\tt arXiv:1312.5742}.
\bibitem[{Hui(2021)}]{Hui:2021tkt}
\bibinfo{author}{Hui, L.} (\bibinfo{year}{2021}).
\newblock \bibinfo{title}{{Wave Dark Matter}}, .
\newblock \href{http://arxiv.org/abs/2101.11735}{\tt arXiv:2101.11735}.
\bibitem[{Hunter(2007)}]{matplotlib}
\bibinfo{author}{Hunter, J.~D.} (\bibinfo{year}{2007}).
\newblock \bibinfo{title}{Matplotlib: A 2d graphics environment}.
\newblock {\it \bibinfo{journal}{Computing in Science \& Engineering}\/},  {\it
  \bibinfo{volume}{9}\/}, \bibinfo{pages}{90--95}.
  \DOIprefix\doi{10.1109/MCSE.2007.55}.
\bibitem[{Ichikawa et~al.(2005)Ichikawa, Fukugita \&
  Kawasaki}]{Ichikawa:2004zi}
\bibinfo{author}{Ichikawa, K.}, \bibinfo{author}{Fukugita, M.}, \&
  \bibinfo{author}{Kawasaki, M.} (\bibinfo{year}{2005}).
\newblock \bibinfo{title}{{Constraining neutrino masses by CMB experiments
  alone}}.
\newblock {\it \bibinfo{journal}{Phys. Rev. D}\/},  {\it
  \bibinfo{volume}{71}\/}, \bibinfo{pages}{043001}.
  \DOIprefix\doi{10.1103/PhysRevD.71.043001}.
  \href{http://arxiv.org/abs/astro-ph/0409768}{\tt arXiv:astro-ph/0409768}.
\bibitem[{Lattanzi \& Gerbino(2018)}]{Lattanzi:2017ubx}
\bibinfo{author}{Lattanzi, M.}, \& \bibinfo{author}{Gerbino, M.}
  (\bibinfo{year}{2018}).
\newblock \bibinfo{title}{{Status of neutrino properties and future prospects -
  Cosmological and astrophysical constraints}}.
\newblock {\it \bibinfo{journal}{Front. in Phys.}\/},  {\it
  \bibinfo{volume}{5}\/}, \bibinfo{pages}{70}.
  \DOIprefix\doi{10.3389/fphy.2017.00070}.
  \href{http://arxiv.org/abs/1712.07109}{\tt arXiv:1712.07109}.
\bibitem[{Lesgourgues(2011{\natexlab{a}})}]{class}
\bibinfo{author}{Lesgourgues, J.} (\bibinfo{year}{2011}{\natexlab{a}}).
\newblock {\it \bibinfo{title}{The Cosmic Linear Anisotropy Solving System
  (CLASS) I: Overview}\/}.
\newblock \bibinfo{type}{Technical Report} CERN.
\newblock \URLprefix \url{[arXiv:1104.2932]}.
\bibitem[{Lesgourgues(2011{\natexlab{b}})}]{lesgourgues2011cosmic}
\bibinfo{author}{Lesgourgues, J.} (\bibinfo{year}{2011}{\natexlab{b}}).
\newblock \bibinfo{title}{The cosmic linear anisotropy solving system (class)
  iii: Comparision with camb for lambdacdm}.
\newblock \href{http://arxiv.org/abs/1104.2934}{\tt arXiv:1104.2934}.
\bibitem[{Lesgourgues et~al.(2013)Lesgourgues, Mangano, Miele \&
  Pastor}]{lesgourgues_mangano_miele_pastor_2013}
\bibinfo{author}{Lesgourgues, J.}, \bibinfo{author}{Mangano, G.},
  \bibinfo{author}{Miele, G.}, \& \bibinfo{author}{Pastor, S.}
  (\bibinfo{year}{2013}).
\newblock {\it \bibinfo{title}{Neutrino Cosmology}\/}.
\newblock \bibinfo{publisher}{Cambridge University Press}.
\newblock \DOIprefix\doi{10.1017/CBO9781139012874}.
\bibitem[{Lesgourgues \& Pastor(2006)}]{Lesgourgues:2006nd}
\bibinfo{author}{Lesgourgues, J.}, \& \bibinfo{author}{Pastor, S.}
  (\bibinfo{year}{2006}).
\newblock \bibinfo{title}{{Massive neutrinos and cosmology}}.
\newblock {\it \bibinfo{journal}{Phys. Rept.}\/},  {\it
  \bibinfo{volume}{429}\/}, \bibinfo{pages}{307--379}.
  \DOIprefix\doi{10.1016/j.physrep.2006.04.001}.
  \href{http://arxiv.org/abs/astro-ph/0603494}{\tt arXiv:astro-ph/0603494}.
\bibitem[{Lesgourgues \& Tram(2011)}]{Lesgourgues_2011}
\bibinfo{author}{Lesgourgues, J.}, \& \bibinfo{author}{Tram, T.}
  (\bibinfo{year}{2011}).
\newblock \bibinfo{title}{The cosmic linear anisotropy solving system (class)
  iv: efficient implementation of non-cold relics}.
\newblock {\it \bibinfo{journal}{Journal of Cosmology and Astroparticle
  Physics}\/},  {\it \bibinfo{volume}{2011}\/}, \bibinfo{pages}{032–032}.
  \URLprefix \url{http://dx.doi.org/10.1088/1475-7516/2011/09/032}.
  \DOIprefix\doi{10.1088/1475-7516/2011/09/032}.
\bibitem[{Lewis et~al.(2000)Lewis, Challinor \& Lasenby}]{camb}
\bibinfo{author}{Lewis, A.}, \bibinfo{author}{Challinor, A.}, \&
  \bibinfo{author}{Lasenby, A.} (\bibinfo{year}{2000}).
\newblock \bibinfo{title}{Efficient computation of cosmic microwave background
  anisotropies in closed friedmann-robertson-walker models}.
\newblock {\it \bibinfo{journal}{The Astrophysical Journal}\/}, . \URLprefix
  \url{[arXiv:astro-ph/9911177]}.
\bibitem[{Ma \& Bertschinger(1995)}]{ma}
\bibinfo{author}{Ma, C.}, \& \bibinfo{author}{Bertschinger, E.}
  (\bibinfo{year}{1995}).
\newblock \bibinfo{title}{Cosmological perturbation theory in the synchronous
  and conformal newtonian gauges}.
\newblock {\it \bibinfo{journal}{Astrophys.J.}\/},  (pp.
  \bibinfo{pages}{7--25}). \URLprefix \url{[arXiv:astro-ph/9506072]}.
\bibitem[{Meurer et~al.(2017)Meurer, Smith, Paprocki, Čertík, Kirpichev,
  Rocklin, Kumar, Ivanov, Moore, Singh, Rathnayake, Vig, Granger, Muller,
  Bonazzi, Gupta, Vats, Johansson, Pedregosa \& Scopatz}]{sympy}
\bibinfo{author}{Meurer, A.}, \bibinfo{author}{Smith, C.},
  \bibinfo{author}{Paprocki, M.}, \bibinfo{author}{Čertík, O.},
  \bibinfo{author}{Kirpichev, S.}, \bibinfo{author}{Rocklin, M.},
  \bibinfo{author}{Kumar, A.}, \bibinfo{author}{Ivanov, S.},
  \bibinfo{author}{Moore, J.}, \bibinfo{author}{Singh, S.},
  \bibinfo{author}{Rathnayake, T.}, \bibinfo{author}{Vig, S.},
  \bibinfo{author}{Granger, B.}, \bibinfo{author}{Muller, R.},
  \bibinfo{author}{Bonazzi, F.}, \bibinfo{author}{Gupta, H.},
  \bibinfo{author}{Vats, S.}, \bibinfo{author}{Johansson, F.},
  \bibinfo{author}{Pedregosa, F.}, \& \bibinfo{author}{Scopatz, A.}
  (\bibinfo{year}{2017}).
\newblock \bibinfo{title}{Sympy: Symbolic computing in python}.
\newblock {\it \bibinfo{journal}{PeerJ Computer Science}\/},  {\it
  \bibinfo{volume}{3}\/}, \bibinfo{pages}{e103}.
  \DOIprefix\doi{10.7717/peerj-cs.103}.
\bibitem[{Nadkarni-Ghosh \& Refregier(2017)}]{Ghosh_2017}
\bibinfo{author}{Nadkarni-Ghosh, S.}, \& \bibinfo{author}{Refregier, A.}
  (\bibinfo{year}{2017}).
\newblock \bibinfo{title}{The einstein–boltzmann equations revisited}.
\newblock {\it \bibinfo{journal}{Monthly Notices of the Royal Astronomical
  Society}\/},  {\it \bibinfo{volume}{471}\/}, \bibinfo{pages}{2391–2430}.
  \URLprefix \url{http://dx.doi.org/10.1093/mnras/stx1662}.
  \DOIprefix\doi{10.1093/mnras/stx1662}.
\bibitem[{Peacock(1997)}]{Peacock_1997}
\bibinfo{author}{Peacock, J.~A.} (\bibinfo{year}{1997}).
\newblock \bibinfo{title}{The evolution of galaxy clustering}.
\newblock {\it \bibinfo{journal}{Monthly Notices of the Royal Astronomical
  Society}\/},  {\it \bibinfo{volume}{284}\/}, \bibinfo{pages}{885–898}.
  \URLprefix \url{http://dx.doi.org/10.1093/mnras/284.4.885}.
  \DOIprefix\doi{10.1093/mnras/284.4.885}.
\bibitem[{Petzold(1983)}]{lsoda}
\bibinfo{author}{Petzold} (\bibinfo{year}{1983}).
\newblock \bibinfo{title}{Automatic selection of methods for solving stiff and
  nonstiff systems of ordinary differential equations}.
\newblock {\it \bibinfo{journal}{SIAM Journal on Scientific and Statistical
  Computing}\/},  {\it \bibinfo{volume}{4}\/}, \bibinfo{pages}{136--148}.
\bibitem[{Pitrou et~al.(2020)Pitrou, Pereira \& Lesgourgues}]{Pitrou:2020lhu}
\bibinfo{author}{Pitrou, C.}, \bibinfo{author}{Pereira, T.~S.}, \&
  \bibinfo{author}{Lesgourgues, J.} (\bibinfo{year}{2020}).
\newblock \bibinfo{title}{{Optimal Boltzmann hierarchies with nonvanishing
  spatial curvature}}.
\newblock {\it \bibinfo{journal}{Phys. Rev. D}\/},  {\it
  \bibinfo{volume}{102}\/}, \bibinfo{pages}{023511}.
  \DOIprefix\doi{10.1103/PhysRevD.102.023511}.
  \href{http://arxiv.org/abs/2005.12119}{\tt arXiv:2005.12119}.
\bibitem[{Poulin et~al.(2019)Poulin, Smith, Karwal \&
  Kamionkowski}]{Poulin:2018cxd}
\bibinfo{author}{Poulin, V.}, \bibinfo{author}{Smith, T.~L.},
  \bibinfo{author}{Karwal, T.}, \& \bibinfo{author}{Kamionkowski, M.}
  (\bibinfo{year}{2019}).
\newblock \bibinfo{title}{{Early Dark Energy Can Resolve The Hubble Tension}}.
\newblock {\it \bibinfo{journal}{Phys. Rev. Lett.}\/},  {\it
  \bibinfo{volume}{122}\/}, \bibinfo{pages}{221301}.
  \DOIprefix\doi{10.1103/PhysRevLett.122.221301}.
  \href{http://arxiv.org/abs/1811.04083}{\tt arXiv:1811.04083}.
\bibitem[{Refregier et~al.(2011)Refregier, Amara, Kitching \& Rassat}]{iCosmo}
\bibinfo{author}{Refregier, A.}, \bibinfo{author}{Amara, A.},
  \bibinfo{author}{Kitching, T.~D.}, \& \bibinfo{author}{Rassat, A.}
  (\bibinfo{year}{2011}).
\newblock \bibinfo{title}{icosmo: an interactive cosmology package}.
\newblock {\it \bibinfo{journal}{Astronomy \& Astrophysics}\/},  {\it
  \bibinfo{volume}{528}\/}, \bibinfo{pages}{A33}. \URLprefix
  \url{http://dx.doi.org/10.1051/0004-6361/200811112}.
  \DOIprefix\doi{10.1051/0004-6361/200811112}.
\bibitem[{Refregier et~al.(2018)Refregier, Gamper, {A. Amara} \&
  Heisenberg}]{pycosmo}
\bibinfo{author}{Refregier, A.}, \bibinfo{author}{Gamper, L.},
  \bibinfo{author}{{A. Amara}}, \& \bibinfo{author}{Heisenberg, L.}
  (\bibinfo{year}{2018}).
\newblock \bibinfo{title}{{PyCosmo}: An integrated cosmological boltzmann
  solver}.
\newblock {\it \bibinfo{journal}{Astronomy and Computing}\/}, . \URLprefix
  \url{[arXiv:1708.05177]}.
\bibitem[{de~Salas et~al.(2021)de~Salas, Forero, Gariazzo, Martínez-Miravé,
  Mena, Ternes, Tórtola \& Valle}]{de_Salas_2021}
\bibinfo{author}{de~Salas, P.~F.}, \bibinfo{author}{Forero, D.~V.},
  \bibinfo{author}{Gariazzo, S.}, \bibinfo{author}{Martínez-Miravé, P.},
  \bibinfo{author}{Mena, O.}, \bibinfo{author}{Ternes, C.~A.},
  \bibinfo{author}{Tórtola, M.}, \& \bibinfo{author}{Valle, J. W.~F.}
  (\bibinfo{year}{2021}).
\newblock \bibinfo{title}{2020 global reassessment of the neutrino oscillation
  picture}.
\newblock {\it \bibinfo{journal}{Journal of High Energy Physics}\/},  {\it
  \bibinfo{volume}{2021}\/}. \URLprefix
  \url{http://dx.doi.org/10.1007/JHEP02(2021)071}.
  \DOIprefix\doi{10.1007/jhep02(2021)071}.
\bibitem[{Schmitt et~al.(2022)Schmitt, Moser, Lorenz \& Refregier}]{uwe}
\bibinfo{author}{Schmitt, U.}, \bibinfo{author}{Moser, B.},
  \bibinfo{author}{Lorenz, C.~S.}, \& \bibinfo{author}{Refregier, A.}
  (\bibinfo{year}{2022}).
\newblock \bibinfo{title}{sympy2c: from symbolic expressions to fast c/c++
  functions and ode solvers in python}.
\newblock \URLprefix \url{https://arxiv.org/abs/2203.11945}.
  \DOIprefix\doi{10.48550/ARXIV.2203.11945}.
\bibitem[{Scolnic et~al.(2018)}]{Scolnic:2017caz}
\bibinfo{author}{Scolnic, D.~M.} et~al. (\bibinfo{year}{2018}).
\newblock \bibinfo{title}{{The Complete Light-curve Sample of Spectroscopically
  Confirmed SNe Ia from Pan-STARRS1 and Cosmological Constraints from the
  Combined Pantheon Sample}}.
\newblock {\it \bibinfo{journal}{Astrophys. J.}\/},  {\it
  \bibinfo{volume}{859}\/}, \bibinfo{pages}{101}.
  \DOIprefix\doi{10.3847/1538-4357/aab9bb}.
  \href{http://arxiv.org/abs/1710.00845}{\tt arXiv:1710.00845}.
\bibitem[{Seljak \& Zaldarriaga(1996)}]{Seljak_1996}
\bibinfo{author}{Seljak, U.}, \& \bibinfo{author}{Zaldarriaga, M.}
  (\bibinfo{year}{1996}).
\newblock \bibinfo{title}{A line-of-sight integration approach to cosmic
  microwave background anisotropies}.
\newblock {\it \bibinfo{journal}{The Astrophysical Journal}\/},  {\it
  \bibinfo{volume}{469}\/}, \bibinfo{pages}{437}. \URLprefix
  \url{http://dx.doi.org/10.1086/177793}. \DOIprefix\doi{10.1086/177793}.
\bibitem[{Tarsitano et~al.(2021)Tarsitano, Schmitt, Refregier, Fluri, Sgier,
  Nicola, Herbel, Amara, Kacprzak \& Heisenberg}]{Tarsitano_2020}
\bibinfo{author}{Tarsitano, F.}, \bibinfo{author}{Schmitt, U.},
  \bibinfo{author}{Refregier, A.}, \bibinfo{author}{Fluri, J.},
  \bibinfo{author}{Sgier, R.}, \bibinfo{author}{Nicola, A.},
  \bibinfo{author}{Herbel, J.}, \bibinfo{author}{Amara, A.},
  \bibinfo{author}{Kacprzak, T.}, \& \bibinfo{author}{Heisenberg, L.}
  (\bibinfo{year}{2021}).
\newblock \bibinfo{title}{Predicting cosmological observables with pycosmo}.
\newblock {\it \bibinfo{journal}{Astronomy and Computing}\/},  {\it
  \bibinfo{volume}{36}\/}, \bibinfo{pages}{100484}. \URLprefix
  \url{https://www.sciencedirect.com/science/article/pii/S221313372100038X}.
  \DOIprefix\doi{https://doi.org/10.1016/j.ascom.2021.100484}.
\bibitem[{Tripathi et~al.(2017)Tripathi, Sangwan \& Jassal}]{Tripathi:2016slv}
\bibinfo{author}{Tripathi, A.}, \bibinfo{author}{Sangwan, A.}, \&
  \bibinfo{author}{Jassal, H.~K.} (\bibinfo{year}{2017}).
\newblock \bibinfo{title}{{Dark energy equation of state parameter and its
  evolution at low redshift}}.
\newblock {\it \bibinfo{journal}{JCAP}\/},  {\it \bibinfo{volume}{06}\/},
  \bibinfo{pages}{012}. \DOIprefix\doi{10.1088/1475-7516/2017/06/012}.
  \href{http://arxiv.org/abs/1611.01899}{\tt arXiv:1611.01899}.
\bibitem[{Virtanen et~al.(2020)Virtanen, Gommers, Oliphant, Haberland, Reddy,
  Cournapeau, Burovski, Peterson, Weckesser, Bright \& et~al.}]{scipy}
\bibinfo{author}{Virtanen, P.}, \bibinfo{author}{Gommers, R.},
  \bibinfo{author}{Oliphant, T.~E.}, \bibinfo{author}{Haberland, M.},
  \bibinfo{author}{Reddy, T.}, \bibinfo{author}{Cournapeau, D.},
  \bibinfo{author}{Burovski, E.}, \bibinfo{author}{Peterson, P.},
  \bibinfo{author}{Weckesser, W.}, \bibinfo{author}{Bright, J.}, \&
  \bibinfo{author}{et~al.} (\bibinfo{year}{2020}).
\newblock \bibinfo{title}{Scipy 1.0: fundamental algorithms for scientific
  computing in python}.
\newblock {\it \bibinfo{journal}{Nature Methods}\/},  {\it
  \bibinfo{volume}{17}\/}, \bibinfo{pages}{261–272}. \URLprefix
  \url{http://dx.doi.org/10.1038/s41592-019-0686-2}.
  \DOIprefix\doi{10.1038/s41592-019-0686-2}.
\bibitem[{van~der Walt et~al.(2011)van~der Walt, Colbert \& Varoquaux}]{numpy}
\bibinfo{author}{van~der Walt, S.}, \bibinfo{author}{Colbert, S.~C.}, \&
  \bibinfo{author}{Varoquaux, G.} (\bibinfo{year}{2011}).
\newblock \bibinfo{title}{The numpy array: A structure for efficient numerical
  computation}.
\newblock {\it \bibinfo{journal}{Computing in Science \& Engineering}\/},  {\it
  \bibinfo{volume}{13}\/}, \bibinfo{pages}{22–30}. \URLprefix
  \url{http://dx.doi.org/10.1109/MCSE.2011.37}.
  \DOIprefix\doi{10.1109/mcse.2011.37}.
\bibitem[{Yoo(2010)}]{Yoo_2010}
\bibinfo{author}{Yoo, J.} (\bibinfo{year}{2010}).
\newblock \bibinfo{title}{General relativistic description of the observed
  galaxy power spectrum: Do we understand what we measure?}
\newblock {\it \bibinfo{journal}{Physical Review D}\/},  {\it
  \bibinfo{volume}{82}\/}. \URLprefix
  \url{http://dx.doi.org/10.1103/PhysRevD.82.083508}.
  \DOIprefix\doi{10.1103/physrevd.82.083508}.
\bibitem[{Yoo et~al.(2009)Yoo, Fitzpatrick \& Zaldarriaga}]{Yoo_2009}
\bibinfo{author}{Yoo, J.}, \bibinfo{author}{Fitzpatrick, A.~L.}, \&
  \bibinfo{author}{Zaldarriaga, M.} (\bibinfo{year}{2009}).
\newblock \bibinfo{title}{New perspective on galaxy clustering as a
  cosmological probe: General relativistic effects}.
\newblock {\it \bibinfo{journal}{Physical Review D}\/},  {\it
  \bibinfo{volume}{80}\/}. \URLprefix
  \url{http://dx.doi.org/10.1103/PhysRevD.80.083514}.
  \DOIprefix\doi{10.1103/physrevd.80.083514}.
\bibitem[{Zumalac\'arregui et~al.(2017)Zumalac\'arregui, Bellini, Sawicki,
  Lesgourgues \& Ferreira}]{Zumalacarregui:2016pph}
\bibinfo{author}{Zumalac\'arregui, M.}, \bibinfo{author}{Bellini, E.},
  \bibinfo{author}{Sawicki, I.}, \bibinfo{author}{Lesgourgues, J.}, \&
  \bibinfo{author}{Ferreira, P.~G.} (\bibinfo{year}{2017}).
\newblock \bibinfo{title}{{hi\_class: Horndeski in the Cosmic Linear Anisotropy
  Solving System}}.
\newblock {\it \bibinfo{journal}{JCAP}\/},  {\it \bibinfo{volume}{08}\/},
  \bibinfo{pages}{019}. \DOIprefix\doi{10.1088/1475-7516/2017/08/019}.
  \href{http://arxiv.org/abs/1605.06102}{\tt arXiv:1605.06102}.

\end{thebibliography}

\end{document}